\documentclass[12pt]{article}
\usepackage{amscd,amssymb,amsmath,latexsym,enumerate}
\usepackage[mathscr]{euscript}
\usepackage{mathrsfs}
\usepackage{epsfig}
\usepackage{fancybox}
\usepackage{verbatim}
\usepackage{tikz}
\usepackage{tikz-cd}
\usepackage{todonotes}
\usepackage{multicol}
\usepackage{graphicx}

\usepackage{color}

\usepackage{xcolor}
\definecolor{MyBlue}{cmyk}{1,0.13,0,0.63}
\definecolor{MyGreen}{cmyk}{0.91,0,0.88,0.52}
\newcommand{\mylinkcolor}{MyBlue}
\newcommand{\mycitecolor}{MyGreen}
\newcommand{\myurlcolor}{black}

\usepackage{hyperref}
\hypersetup{%
  bookmarksnumbered=true,bookmarksopen=false,%
  plainpages=false,
  linktocpage=true,%
  colorlinks=true,breaklinks=true,%
  linkcolor=\mylinkcolor,citecolor=\mycitecolor,urlcolor=\myurlcolor,%
  pdfpagelayout=OneColumn,%
  pageanchor=true,%
}

\title{Callias-type operators associated to spectral triples}

\author{Hermann Schulz-Baldes and Tom Stoiber
\\
\\
{\small Department Mathematik,  Friedrich-Alexander-Universit\"at Erlangen-N\"urnberg}
\\
{\small Cauerstr. 11, D-91058 Erlangen, Germany}
\\
{\small Email: schuba@mi.uni-erlangen.de, tom.stoiber@fau.de}
}

\date{ }

\newtheorem{theorem}{Theorem}
\newtheorem{proposition}[theorem]{Proposition}
\newtheorem{lemma}[theorem]{Lemma}
\newtheorem{corollary}[theorem]{Corollary}
\newtheorem{definition}[theorem]{Definition}

\newcommand{\CM}{{\mathbb C}}
\newcommand{\NM}{{\mathbb N}}
\newcommand{\RM}{{\mathbb R}}
\newcommand{\SM}{{\mathbb S}}

\newcommand{\ZM}{{\mathbb Z}}

\newcommand{\KM}{{\mathbb K}}

\newcommand{\Aa}{{\cal A}}
\newcommand{\Ee}{{\cal E}}

\newcommand{\Bb}{{\cal B}}
\newcommand{\Dd}{{\cal D}}
\newcommand{\Ff}{{\cal F}}

\newcommand{\Ss}{{\cal S}}
\newcommand{\Oo}{{\cal O}}
\newcommand{\Tt}{{\cal T}}

\newcommand{\Nn}{{\cal N}}
\newcommand{\Mm}{{\cal M}}
\newcommand{\Cc}{{\cal C}}

\newcommand{\Kk}{{\cal K}}
\newcommand{\Hh}{{\cal H}}

\newcommand{\ec}{{\rm ec}}

\providecommand{\abs}[1]{\left \lvert#1 \right \rvert} 
\providecommand{\norm}[1]{\left \lVert#1 \right \rVert}
\newcommand{\Fsa}{\mathcal{F}_\textup{sa}}
\newcommand{\tInd}{\Tt\mbox{-}\textup{Ind}}
\newcommand{\one}{{\bf 1}}

\newcommand{\Tr}{\mbox{\rm Tr}}

\newcommand{\SF}{{\rm Sf}}

\newcommand{\Ind}{{\rm Ind}} 
\newcommand{\Ker}{{\rm Ker}}

\newcommand{\sgn}{{\rm sgn}} 
 
\newcommand{\diag}{{\rm diag}}

\newcommand{\Diracphase}{F}

\begin{document}

\maketitle

\begin{abstract}
Callias-type (or Dirac-Schr\"odinger) operators associated to abstract semifinite spectral triples are introduced  and their indices are computed in terms of an associated index pairing derived from the spectral triple. The result is then interpreted as an index theorem for a non-commutative analogue of spectral flow. Both even and odd spectral triples are considered, and both commutative and non-commutative examples are given.




\end{abstract}



\section{Outline}
\label{sec-outline}


The Callias index theorem \cite{Cal,Ang2,GW} and its even dimensional analogue \cite{GH,Bunke,FH} give formulas for the index of Dirac operators on non-compact manifolds which are perturbed by self-adjoint potentials that act on sections of a finite-dimensional vector bundle and are invertible at infinity. There are many possible generalizations, for example, one can allow infinite-dimensional vector bundles as in the Robbin-Salomon theorem \cite{RoS} or Hilbert-module bundles of finite type \cite{C,BC}. This paper generalizes in a different direction, namely the underlying manifold is replaced by a semifinite spectral triple and the perturbing potentials will be elements of a certain multiplier algebra. In this abstract setting one can still express the index of a Callias-type operator in terms of an index pairing derived from the spectral triple. Concretely, if $(\Nn, D, \mathscr{A})$ is a semifinite spectral triple with trace $\Tt$ and a Callias potential $H$ is a self-adjoint differentiable multiplier of $\Aa=C^*(\mathscr{A})$ which is invertible modulo $\Aa$ (in the sense of Definition~\ref{def-AsympInv} below), then the Callias-type operator $\kappa D + \imath H$ is $\Tt$-Fredholm for small enough $\kappa>0$ and
$$
\Tt{\mbox{-}}\Ind(\kappa D + \imath H) \;=\; \langle [U]_1, [D]\rangle
\;,
$$
where $U = \exp(\imath \pi(G(H)+\one))$ is a unitary defining a $K$-theory class in $K_1(\Aa)$ for $G$ a suitable switch function, and $\imath=\sqrt{-1}$. The precise statement is given in Section~\ref{sec-Main}, for the case of unbounded $H$ in Section~\ref{sec-Unbounded}. Furthermore, Section~\ref{sec-Even} states and proves an even analogue  for pairings of an even spectral triple with a potential $H$ having a further symmetry. Section~\ref{sec-Unbounded} then also covers the unbounded even case. Section~\ref{sec-Examples} presents classical and new examples.

\vspace{.2cm}

In the commutative setting one would take for $\mathscr{A}$ the algebra of smooth compactly supported fiberwise compact multiplication operators on sections $\Hh$ of a vector bundle over a complete Riemannian manifold $X$, for $D$ a weakly elliptic first-order differential operator and $\Tt=\Tr_{L^2(X)}\otimes\Tr_\Hh$. Results comparable to our main result in that commutative setting with infinite-dimensional vector bundles have been obtained  by Kaad and Lesch \cite{KL} and more recently also with less regularity assumptions by van den Dungen \cite{vD}. These proofs rely on previous work on unbounded Kasparov products. In essence the strategy in \cite{KL,vD} is to prove that a Callias-type operator is an unbounded representative of the product of a $K$-homology class defined by a first-order differential operator and a class in the $K_1$-group over the continuous functions on the manifold defined by a self-adjoint multiplication operator. As discussed in Section~\ref{sec-CompareKK} this approach can also be made to work in the present more general noncommutative setting with some technical limitations. Instead, here we provide a new and rather elementary proof using semifinite spectral flow and explicit operator homotopies. Even for the classical case ({\it e.g.} \cite{GW,GH}), the argument constitutes a considerable simplification of the proof.

\vspace{.2cm}

In the special case where $X$ is the real line the potential $H$ represents a path of self-adjoint Fredholm operators and the index coincides with the spectral flow of the family. This analogy has been carried further by \cite{KL,vD} who call the index pairing $\langle [U]_1, [D]\rangle$ a spectral flow also in higher dimensions, motivated by notions from $KK$-theory. We follow that interpretation and therefore consider the index pairing as a non-commutative analogy of spectral flow.

\vspace{.2cm}

\section{Callias-type operators with bounded potentials}

Let $\Nn$ be a von Neumann algebra with semifinite normal faithful trace $\Tt$ acting on a Hilbert space $\Hh$. For the convenience of the reader, several facts about the trace $\Tt$, the set $\Kk_\Tt$ of $\Tt$-compact operators and the notion of Breuer-Fredholm or $\Tt$-Fredholm operators and their index $\Tt\mbox{-}\Ind$ are recalled in Appendix~\ref{app-SemiInd}. 
An (odd) semifinite spectral triple $(\Nn, D,\mathscr{A})$ \cite{CareyEtAl,CGPR} consists of an unbounded self-adjoint operator $D$ affiliated with $\Nn$ and a $*$-algebra $\mathscr{A} \subset \Nn$ such that 
\begin{enumerate}
\item[(i)] Each $A\in \mathscr{A}$ preserves the domain of $D$ and the hence densely defined operator $[D,A]$ extends to a bounded element of $\Nn$.
\item[(ii)] For each $A\in \mathscr{A}$ the product $A(1+D^2)^{-\frac{1}{2}} $ is $\Tt$-compact.
\end{enumerate}

Let $\Aa=C^*(\mathscr{A})$ be the $C^*$-algebra generated by $\mathscr{A}$. By default a spectral triple produces an index pairing with the $K$-theory group $K_1(\Aa)$ through \cite{CareyEtAl}
\begin{equation} 
\label{eq-IndexPairing}
\langle [U]_1, [D]\rangle \;=\; \Tt\mbox{-}\Ind(PUP + \one - P)\; \in\; \RM
\;,
\end{equation}
where $P=\chi(D>0)$ and the $\Tt$-index and the operator on the r.h.s. is $\Tt$-Fredholm for any representative $U \in \one + \mathscr{A}$ defining a class in $K_1(\Aa)$. The potentials for our Callias-type operators will be recruited from a larger algebra:

\begin{definition} 
\label{def-CalliasType}
Let $(\Nn, D,\mathscr{A})$ be a semifinite spectral triple.
\begin{enumerate}[{\rm (i)}]

\item The multiplier algebra $M(\Aa, \Nn)$ is the idealizer of $\Aa$ in $\Nn$, {\it i.e.} the largest $C^*$-subalgebra of $\Nn$ such that $M(\Aa, \Nn)\Aa \subset \Aa$ and $\Aa M(\Aa, \Nn) \subset \Aa$. Elements of $M(\Aa, \Nn)$ are also called $\Aa$-multipliers. 

\item An $\Aa$-multiplier $H\in M(\Aa, \Nn)$ is differentiable w.r.t. $D$ if $H$ preserves $\mathrm{Dom}(D)$ and $[D,H]$ extends to a bounded operator.

\item For a self-adjoint differentiable $\Aa$-multiplier $H$ the associated Callias-type operator on the domain $\mathrm{Dom}(D)$ is defined by
$$
D_{\kappa, H} \;=\; \kappa D + \imath H
\;,
\qquad
\kappa\,>\,0
\;.
$$
\end{enumerate}

\end{definition}

The parameter $\kappa$ can be interpreted as the scale of the noncommutative space quanta. It plays a prominent role in the following. 
It will next be useful to pass to a self-adjoint supersymmetric operator $L_{\kappa,H}$ which, due to the prior works \cite{LS1,LS2,ST2021}, will also be referred to as the spectral localizer. 

\begin{proposition}
\label{prop-KatoRellich}
For a  self-adjoint $\Aa$-multiplier $H$, the adjoint of the Callias-type operator $D_{\kappa, H}$ is $(D_{\kappa, H})^*=\kappa D - \imath H $ and the spectral localizer
$$
L_{\kappa,H} 
\;=\; 
\begin{pmatrix}
0 & D_{\kappa, H}^*\\ D_{\kappa, H} & 0
\end{pmatrix} 
$$
is self-adjoint on the domain $\mathrm{Dom}(D)^{\times 2}$. Moreover, $D_{\kappa, H}$ and $L_{\kappa,H}$ are affiliated to $\Nn$ and $M_2(\Nn)$ respectively. 
\end{proposition} 

\noindent {\bf Proof.}
Note that
$$
L_{\kappa,H} 
\;=\; 
\kappa \,
\begin{pmatrix}
0 &  D  \\  D  & 0
\end{pmatrix} 
\;+\;
\begin{pmatrix}
0 &  - \imath H \\  \imath H & 0
\end{pmatrix} 
\;.
$$
The first summand is self-adjoint, and the second is a bounded self-adjoint perturbation leaving the domain $\mathrm{Dom}(D)^{\times 2}$ invariant. Hence the self-adjointness of $L_{\kappa,H}$ immediately follows from the Kato-Rellich theorem and this also implies that $(D_{\kappa, H})^*=D_{\kappa, -H}$. As to the last claim,  recall that an equivalent condition for the affiliation of an operator $T$ is that each unitary $U\in \Nn'$ preserves the domain of $T$ and commutes $UTU^*=T$ (see {\it e.g.} \cite{KadisonRingrose}).
\hfill $\Box$

\vspace{.2cm}

The following often deals with operators in or affiliated to the matrix algebras $M_2(\Nn)$ and $M_2(\Aa)$. The former is supplied with the natural trace $\Tt \circ \Tr$, but for notational convenience we will denote it by the same letter and speak of $\Tt$-compact and $\Tt$-Fredholm operators with no regard for the size of the matrices.

\vspace{.2cm}

The following provides a criterion for Callias-type operators to be $\Tt$-Fredholm:

\begin{definition}
\label{def-AsympInv}
A self-adjoint $\Aa$-multiplier $H$ is called asymptotically invertible w.r.t. $\Aa$ if there is a positive element $V \in \Aa$ such that $H^2+V>0$ is invertible. An asymptotically invertible and differentiable self-adjoint $\Aa$-multiplier $H$ is called a Callias potential. 
\end{definition}

In the classical case of a Riemannian manifold $X$ where $\Aa=C_0(X,\Kk(\Hh))$ and $H$ is given by an operator-valued bounded function $x\in X\mapsto H_x\in\Bb(\Hh)$, the asymptotic invertibility is indeed equivalent to the uniform invertibility of $H_x$ outside a compact subset $K\subset X$, namely there is a positive constant $c$ such that $H_x^2\geq c\,\one$ for $x\in X\setminus K$. Proposition~\ref{prop-SmallSupp} shows that $V$ in Definition~\ref{def-AsympInv} can always be chosen as a spectral function of $H$ itself. This also implies that asymptotic invertibility of $H$ is equivalent to the invertibility of $\pi(H)$ in the quotient algebra $M(\Aa,\Nn)/\Aa$.

\begin{proposition}
\label{prop:fredholm1}
If $H$ is a Callias potential, then there exists a $\kappa_0>0$ such that $L_{\kappa,H}$ and therefore $D_{\kappa,H}$, $D^*_{\kappa,H}$ are $\Tt$-Fredholm for all $\kappa\in(0,\kappa_0]$. Moreover, the $\Tt$-index of the Callias-type operator given by
$$
\Tt\mbox{-}\Ind(D_{\kappa,H}) \;=\; \Tt(\Ker(D_{\kappa,H})) - \Tt(\Ker(D^*_{\kappa,H})) \, \in \RM
$$
is independent of $\kappa\in(0,\kappa_0]$.
\end{proposition}

One may also view $\Tt\mbox{-}\Ind(D_{\kappa,H})$ as the supersymmetric index of $L_{\kappa,H}$. The proof of Proposition~\ref{prop:fredholm1} will use smooth functional calculus of a self-adjoint operator via the well-known Helffer-Sj\"ostrand or Dynkin formula \cite{Davies95}. For later use in Section~\ref{sec-Unbounded}, let us recall the details for a possibly unbounded self-adjoint operator $H$.  For $\rho \in \RM$, let $\Ss^{\rho}(\RM)$ denote the set of smooth functions   $f:\RM\to\RM$ satisfying
$$
\lvert\partial^k f(x)\rvert \;\leq\; C_k (1+x^2)^{\frac{\rho-k}{2}}
\;, \qquad k\in \NM\;.
$$
Then there exists for any $N>0$ an almost analytic representation $\tilde{f}_N$ of $f$ supported in a complex set $G$ of the form $G=\{x+\imath y: \,\lvert y\rvert < 2\sqrt{1+x^2}\}$ which coincides with $f$ on $\RM$ and 
\begin{equation}
	\label{eq:hsbounds}
\lvert \partial_{\overline{z}} \tilde{f}(z) \rvert 
\;\leq\; c_N \tilde{C}_{N+1}  (1+x^2)^{\frac{\rho-1-N}{2}}\, \lvert\Im m (z)\rvert^{N}
\end{equation}
for a universal constant $c_N$ and $\tilde{C}_{N+1}=\sum_{k=1}^{N+1} C_k$, (see {\it e.g.} \cite[Lemma 2.2.1]{Davies95}).
Provided $\rho < 0$, the Helffer-Sj\"ostrand representation 
\begin{equation}
\label{eq:hsrep}
f(H) \;=\; \int_G (\partial_{\overline{z}} \tilde{f}_N(z)) (H-z)^{-1} \mathrm{d}z \wedge \mathrm{d}\overline{z}
\end{equation}
is a norm-convergent integral for any $N\geq 1$. For a complex-valued function $f:\RM\to\CM$ this can be done for real and imaginary part separately.

\vspace{.2cm}

\noindent{\bf Proof} of Proposition~\ref{prop:fredholm1}: Let $\chi:[0,\infty)\to[0,1]$ be a smooth function with $\chi(0)=1$ and vanishing outside $[0,\frac{g^2}{2}]$ We have to show that $\chi(L_{\kappa,H}^2)$ is $\Tt$-compact for which  formula \eqref{eq:hsrep} is used with a quasianalytic extension $\tilde{\chi}_N$ of $\chi$. To control the resolvent of $L_{\kappa,H}^2$, let us note that
\begin{align*}
L_{\kappa,H}^2
&
\;=\;
\begin{pmatrix}
\kappa^2\,D^2+H^2 & 0 \\ 0 & \kappa^2\,D^2+H^2
\end{pmatrix}
\;+\;
\imath\kappa
\begin{pmatrix}
[D,H] & 0 \\ 0 & -[D,H]
\end{pmatrix}
\;.
\end{align*}
Now let $V$ and $g>0$ be such that $H^2+V\geq g^2\one$. Then with $\tilde{V}=V\oplus V$
\begin{equation}
\label{eq-kappa0bound}
L_{\kappa,H}^2+\tilde{V}
\;\geq\;
\big(g^2
\,-\,
\kappa
\|[D,H]\|\big)\one_2
\;,
\end{equation}
which shows that $L_{\kappa,H}^2+\tilde{V}$ is invertible for $\kappa$ sufficiently small. Now replace the resolvent identity into the Helffer-Sj\"ostrand  formula
$$
\chi(L_{\kappa,H}^2) 
\;=\; 
\int_G (\partial_{\overline{z}} \tilde{\chi}_N(z)) 
\left[
(L_{\kappa,H}^2+\tilde{V}-z)^{-1}
\,+\,
(L_{\kappa,H}^2+\tilde{V}-z)^{-1}\tilde{V}(L_{\kappa,H}^2-z)^{-1}
\right]
\mathrm{d}z \wedge \mathrm{d}\overline{z}
\;.
$$
The first summand is $\chi(L^2_{\kappa,H}+\tilde{V})$ and hence vanishes if $\kappa \leq \kappa_0$ where $\kappa_0 = \frac{g^2}{2}  \norm{[D,H]}^{-1}$. For the remaining term, applying the resolvent identity again shows
$$
\tilde{V}(L_{\kappa,H}^2-z)^{-1}
\;=\;
\tilde{V}(\kappa^2D^2\otimes\one_2-z)^{-1}
\big[
\one+(H^2\otimes\one_2+\imath\kappa [D,H]\otimes\sigma_3(L_{\kappa,H}^2-z)^{-1}
\big]
\;.
$$
As $\tilde{V}\in M_2(\Aa)$, the factor $\tilde{V}(\kappa^2D^2\otimes\one_2-z)^{-1}$ is $\Tt$-compact due to the definition of the spectral triple. As $\Kk_\Tt$ is a norm closed ideal and the integral in the Helffer-Sj\"ostrand  formula is norm convergent, this implies that $\chi(L_{\kappa,H}^2)$ is $\Tt$-compact.

\vspace{.1cm}

The final claim follows from the fact that $\kappa\mapsto  D_{\kappa,H}$ is continuous in the gap topology so that the $\Tt$-index is constant along this path.
\hfill $\Box$

\vspace{.2cm}

The next result is the last technical preparation. 

\begin{proposition}
\label{prop-SmallSupp}
Let $H$ be a self-adjoint $\Aa$-multiplier satisfying $H^2+V> g^2\one$ for some $V=V^*\in\Aa$. Then for every function $f:\RM\to\CM$ supported by $[-\frac{g}{2},\frac{g}{2}]$ one has $f(H)\in\Aa$.
\end{proposition}

\noindent{\bf Proof.} Let $\chi:[0,\infty)\to [0,1]$ be a smooth function satisfying $\chi(\lambda)=1$ for $\lambda\leq\frac{g^2}{4}$ and $\chi(\lambda)=0$ for $\lambda\geq g^2$. Then $\chi(H^2+V)=0$ by hypothesis so that the Helffer-Sj\"ostrand formula and resolvent identity imply
$$
\chi(H^2)
\;=\;
\int_G (\partial_{\overline{z}} \tilde{\chi}_N(z)) 
\,(H^2+V-z)^{-1}V(H^2-z)^{-1}\,
\mathrm{d}z \wedge \mathrm{d}\overline{z}
\;.
$$
As $V\in\Aa$, this implies $\chi(H^2)\in\Aa$ because $\Aa$ is a norm-closed ideal in $M(\Aa,\Nn)$. Now by construction, $f(H)=\chi(H^2)f(H)$ and therefore invoking the ideal property once again leads to  $f(H)\in\Aa$.
\hfill $\Box$

\section{Main result for bounded Callias potentials}
\label{sec-Main}

\begin{proposition}
\label{prop-ExpF}
Let $H$ be a self-adjoint $\Aa$-multiplier which is asymptotically invertible w.r.t. $\Aa$ and satisfies $H^2+V > g^2\one$ for some $g>0$ and self-adjoint $V\in \Aa$. Let $G: \RM \to \RM$ be a smooth nondecreasing odd function taking values $-1$ below $-\frac{g}{2}$ and $1$ above $\frac{g}{2}$. Then the unitary operator 
$$
U \;=\; e^{\imath\pi  (G(H)+\one)}
$$ 
defines a class in $K_1(\Aa)$ which does not depend on the function $G$. It represents the image of the spectral projection $[\chi(H + \Aa < 0)]_0 \in K_0(M(\Aa,\Nn)/\Aa)$ under the exponential map in $K$-theory $\partial_0: K_0(M(\Aa, \Nn)/\Aa)\to K_1(\Aa)$ associated to the short exact sequence
\begin{equation}
\label{eq-ExactSeq}
0\;\to\;\Aa\;\to\;M(\Aa,\Nn)\;\to M(\Aa,\Nn)/\Aa\;\to\;0
\;,
\end{equation}
namely
$$
[U]_1 \;=\; \partial_0 [\chi(H + \Aa < 0)]_0
$$
\end{proposition}

\noindent{\bf Proof.}
By construction $e^{\imath\pi  (G(\lambda)+1)}-1$ is supported in $[-\frac{g}{2},\frac{g}{2}]$ and one therefore has $U -\one\in \Aa$ by Proposition~\ref{prop-SmallSupp}. Naturally it defines the class $[U]_1$ in $K_1(\Aa)$. The second claim results from the fact that $\frac{1}{2}(G(H)+\one)$ is a lift of the projection $\one-\chi(H + \Aa < 0)$ into $M(\Aa,\Nn)$.
\hfill $\Box$

\vspace{.2cm}

Clearly one can also replace $M(\Aa,\Nn)$ by any smaller $C^*$-algebra that contains $H$ and has $\Aa$ as an ideal, since by naturality all computations pull back to the connecting map of $0 \to \Aa \to C^*(H,\Aa) \to C^*(H,\Aa)/\Aa\to 0$ where $C^*(H,\Aa)=C^*(H) + \Aa$.

\begin{definition}
\label{def-SFodd}
For an asymptotically invertible $\Aa$-multiplier $H$, the $D$-spectral flow is defined as an index pairing in the sense of \eqref{eq-IndexPairing} by
$$
\SF_D(H) \;=\; \langle [e^{\imath\pi  (G(H)+\one)}]_1, [D]\rangle \; \in \; \RM
$$
for any admissible function $G$ as specified in {\rm Proposition~\ref{prop-ExpF}} above.
\end{definition}

Let us briefly justify why this definition applied to a particular set-up indeed reduces to the standard notion of semifinite spectral flow. Let  $(\mathfrak{n},\tau)$ be a semifinite von Neumann algebra and $\mathfrak{n}_\Tt$ the traceclass elements. Then a differentiable path $x\in \RM\mapsto H_x\in \mathfrak{n}$ of self-adjoint Fredholm operators with invertible limits can be paired with a winding number $1$-cocycle to give the spectral flow in the formulation of Wahl \cite{Wahl2008}, see Definition~\ref{def-SF} in the appendix where this is spelled out for a finite interval. This latter spectral flow coincides with Definition~\ref{def-SFodd} if one chooses 
$$
(\mathscr{A},\Nn,\Tt,D)
\;=\;
\Big(C^\infty_c(\RM,\mathfrak{n}_\Tt), L^\infty(\RM,\mathfrak{n}),\Tt=\int dx\otimes \tau,\imath \partial_x\otimes\one\Big)
\;.
$$
The classical case is obtained when $ \mathfrak{n}=\Bb(\Hh)$ and $\tau=\Tr$. More generally, for differentiable families $x\in \RM^d\mapsto H_x\in \mathfrak{n}$ with odd $d$, the above defintion reduces to a volume integral version of a generalized multiparameter spectral flow, see the discussion in Section~\ref{sec-ClassCall}. Definition~\ref{def-SFodd} further conceptualizes these special cases, and is in the spirit of \cite[Definition~8.9]{KL} and \cite[Definition 2.18]{vD} where a $K$-theory valued spectral flow is introduced. The following main result shows that the index of a Callias-type operator is equal to the spectral flow in the sense of Definition~\ref{def-SFodd}:

\begin{theorem}
\label{thm-Main}
Let $H$ be a Callias potential for the semifinite spectral triple $(\Nn, D, \mathscr{A})$. Set
\begin{equation}
\label{eq-kappaCond}
\kappa_0
\;=\;
\frac{g^2}{2\,\|[D,H]\|}
\;,
\qquad
g^2\;=\;
\min\,\sigma(H^2+\Aa)
\;.
\end{equation}
Then for all $\kappa\in(0,\kappa_0)$ 
$$
\Tt\mbox{-}\Ind(D_{\kappa, H}) \;=\; \SF_D(H)
\;.
$$
\end{theorem}

Let us comment that the equality of index and spectral flow in general does not hold for large values of $\kappa$. Indeed, a  counterexample (with $X=\RM$ and $\Aa=C_0(X,\Kk(\Hh))$ with an infinite dimensional fiber Hilbert space $\Hh$) can be found in the work of Abbondandolo and Majer \cite[Section~7]{AM}. Theorem~\ref{thm-Main} only concerns the semiclassical regime of small $\kappa$, or otherwise stated the limiting index for small $\kappa$. For unbounded $H$ the situation may be different. Indeed, the Robbin-Salamon theorem  states that for one-dimensional potentials growing at infinity, all values of $\kappa$ are allowed. This case is covered by Theorem~\ref{thm-MainUnbounded} below.

\section{Proof of the main result}
\label{sec-ProofOdd}

The first step is make the Dirac operator invertible which can be achieved by a standard doubling trick. More precisely, set
\begin{equation}
\label{eq:doubling}
\tilde{D} \;=\; \begin{pmatrix}
D & \mu \\  \mu & -D
\end{pmatrix}, 
\qquad 
\tilde{H} \;=\; \begin{pmatrix}
H & 0 \\  0 & \one
\end{pmatrix}
\end{equation}
for some $\mu > 0$ and also define 
$$
\tilde{D}_{\kappa,\tilde{H}} \;=\; \kappa \tilde{D} + \imath \tilde{H},
\qquad \tilde{L}_{ \kappa,\tilde{H}}\;=\;\begin{pmatrix}
0 & \tilde{D}_{\kappa,\tilde{H}}^* \\ \tilde{D}_{\kappa,\tilde{H}} & 0
\end{pmatrix}.$$
Self-adjointness of $\tilde{L}_{ \kappa,\tilde{H}}$ follows again from the Kato-Rellich theorem, but the Fredholm property can depend nontrivially on $\kappa$ and $\mu$. In Lemma~\ref{lemma:fredholm_localizer} below, we show that one may choose $\mu=\Oo(1)$ and then $\kappa\leq\Oo(\mu)$. For $\mu=0$ the additivity of the Fredholm index gives
\begin{equation}
\label{eq:indtilde}
\Tt\mbox{-}\Ind(D_{\kappa,H})\;=\;\Tt\mbox{-}\Ind(\tilde{D}_{\kappa,\tilde{H}})
\end{equation} 
and then the index stays unchanged for non-vanishing $\mu$ as long as the Fredholm-property is not violated. It is sufficient to prove the index formula for some sufficiently small $\kappa$ because it then immediately holds for all $\kappa$ as stated in Theorem~\ref{thm-Main}. The next step is to express the index pairing as a spectral flow and to separate $\tilde{D}$ and $\tilde{H}$ in the $2\times 2$ matrix.

\begin{lemma}
\label{lem-SFInd}
For any $m>0$ and $\kappa$ sufficiently small, the $\Tt$-index of $\tilde{D}_{\kappa,\tilde{H}}$ can be computed as spectral flow along a straight-line path:
\begin{align*}
\Tt\mbox{-}\Ind(\tilde{D}_{\kappa,\tilde{H}})\;=\; -\, \SF(\begin{pmatrix}
\kappa \tilde{D} &  \tilde{H}-\imath m \\  \tilde{H}+\imath m & - \kappa \tilde{D}
\end{pmatrix},\begin{pmatrix}
\kappa \tilde{D} & \tilde{H}+\imath m \\ \tilde{H}- \imath m & - \kappa \tilde{D}
\end{pmatrix}).
\end{align*}
\end{lemma}
\noindent{\bf Proof.}
By conjugation with the unitary matrix $C =\frac{1}{\sqrt{2}} \begin{pmatrix}
1 & \imath  \\ 1& -\imath
\end{pmatrix}$, one has
\begin{align*}
&\SF(\begin{pmatrix}
\kappa \tilde{D} & \tilde{H} -\imath m\\ \tilde{H}+\imath m & - \kappa \tilde{D}
\end{pmatrix},\begin{pmatrix}
\kappa \tilde{D} & \tilde{H}+\imath m \\ \tilde{H}-\imath m  & - \kappa \tilde{D}
\end{pmatrix}) \\
&\;=\; \SF(\begin{pmatrix}
-m & \kappa \tilde{D}+\imath \tilde{H} \\ \kappa \tilde{D}-\imath \tilde{H} & m
\end{pmatrix},\begin{pmatrix}
m & \kappa \tilde{D}+\imath \tilde{H} \\ \kappa \tilde{D}-\imath \tilde{H} & - m
\end{pmatrix})
\end{align*}
and so the claim follows from Proposition~\ref{prop:indexspectralflow}.
\hfill $\Box$

\vspace{.2cm}


The next step will be to deform the off-diagonal entries (more precisely the path in the lower left corner from $\tilde{H}+\imath m$ to $\tilde{H}-\imath m$) in the matrices on the r.h.s. of Lemma~\ref{lem-SFInd} into a unitary, without changing the spectral flow.  This will be done by functional calculus in $\tilde{H}$ (so for every spectral value $\lambda\in\RM$) by homotopically deforming the function 
\begin{equation}
\label{eq-Path0}
t \in [0,1] \;\mapsto\; f_{t,0}(\lambda) \;=\; (1-t) (\lambda + \imath m)+ t(\lambda - \imath m)
\end{equation}
into
\begin{equation}
\label{eq-Path1}
t \in [0,1]\; \mapsto \;f_{t,1}(\lambda) \;=\; (1-t)  \imath e^{-\imath \frac{\pi}{2}G(\lambda)}+t (-\imath) e^{\imath \frac{\pi}{2}G(\lambda)}
\;.
\end{equation}
The second path is constructed to contain a square root of the unitary $U=e^{\imath \pi (G({H})+\one)}$ appearing in the image of the exponential map in Proposition~\ref{prop-ExpF}. The main analytical difficulty that has to be addressed next is that along such a deformation the Fredholm property has to be maintained. For this purpose, let use the odd spectral localizer
$$L^o_{\kappa, f}\;=\;\begin{pmatrix}
\kappa \tilde{D} & f(\tilde{H})^* \\ f(\tilde{H}) & -\kappa \tilde{D} 
\end{pmatrix} $$
for an arbitrary differentiable function $f:\RM\to\CM$. The Fredholm property of $L^o_{\kappa,f }$ can again be checked by formally squaring
$$(L^o_{\kappa, f })^2 \;=\; \begin{pmatrix}
\kappa \tilde{D}^2+|f(\tilde{H})|^2 & \kappa[\tilde{D},f(\tilde{H})^*] \\ \kappa[f(\tilde{H}) , \tilde{D}]  & \kappa^2 \tilde{D}^2 +|f(\tilde{H})|^2
\end{pmatrix} 
\;,
$$
with modified commutator by the doubling given by 
$$[f(\tilde{H}), \tilde{D}] \;=\; \begin{pmatrix}
[f(H), D] & \mu (f(\one)-f(H)) \\ - \mu(f(\one)-f(H)) & 0
\end{pmatrix}
$$
For the control of $[f(H), D]$, let us recall: 

\begin{lemma}
\label{lem-BR}
For every smooth function $f$, the commutator $[D,f(H)]$ extends from $\mathrm{Dom}(D)$ to a bounded operator with norm bound
$$
\|[D, f(H)] \| \;\leq\; \tilde{C}_3 \lVert H \rVert\, \| [D, H]\|
$$
where $\tilde{C}_3=C\sum_{i=0}^3\|f^{(i)}\|_\infty$ is a constant.
\end{lemma}

\noindent {\bf Proof.} Using the Helffer-Sj\"ostrand formula \eqref{eq:hsrep} for $N=2$, one has the norm convergent integral 
$$
[D,f(H)] \;=\; -\int_G (\partial_{\overline{z}} f_2(z)) (H-z)^{-1}[D,H](H-z)^{-1} \mathrm{d}z \wedge \mathrm{d}\overline{z}.
$$
and the claimed bound on the commutator follows immediately. An alternative proof can also be given using
$\|[D,f(H)]\|\leq\|(\Ff f')\|_{L^1(\RM)} \|[D,H]\|$ where $\Ff$ is the Fourier transform  \cite[Lemma 10.15]{GVF}.
\hfill $\Box$

\vspace{.2cm}

Using Lemma~\ref{lem-BR}, one obtains
\begin{align}
(L^o_{\kappa, f })^2 
&
\;\geq\; 
\kappa^2 \mu^2 + |f(\tilde{H})|^2 - \kappa\|[f(\tilde{H}), \tilde{D}]\|
\nonumber
\\
&
\;\geq\; 
\kappa^2 \mu^2 + |f(\tilde{H})|^2 - \kappa \big(\tilde{C}_3 \lVert H \rVert\, \|[D,H]\|+\mu\|f(H)-f(\one)\|  \big).
\label{eq-LowBound}
\end{align}
From this, we will now need to derive a quantitative lower bound on the essental spectrum of $(L^o_{\kappa, f })^2 $, namely a lower bound on $(L^o_{\kappa, f })^2 + M_2(\Kk_\Tt)$. For that purpose and the remainder of the section let us now assume that $\lvert H \rvert > \one \mod \Aa$ which can be achieved without loss of generality by rescaling $H$ and all other parameters ($\kappa$, $\mu$, $m$, etc.). 

\begin{lemma}
\label{lemma:fredholm_localizer}
Associated to a smooth function $f\in C(\sigma(\tilde{H}),\CM)$, let $\tilde{C}_3$ be as in Lemma~\ref{lem-BR} and set 
$$
c_1 \;=\;\min_{|\lambda| \geq 1} |f(\lambda)|, \qquad c_2 \;=\; 2\,\|f\|_\infty, 
\;,
$$ and $\kappa$ such that
\begin{equation}
\label{eq:lowerbound}
\tfrac{1}{2} c_1^2 + \mu^2 \kappa^2 - \kappa (\tilde{C}_3 \lVert H \rVert \| [D, H]\| + \mu c_2) \;>\; 0
\;,
\end{equation}
then $(L^o_{\kappa, f })^2$ is a self-adjoint $\Tt$-Fredholm operator with spectrum mod $M_2(\Kk_{\Tt})$ bounded from below by $\frac{1}{2}c_1^2 + \mu^2 \kappa^2 - \kappa (C c_3 \lVert H \rVert \| [D, H]\| + \mu c_2)$.
\end{lemma}

\noindent{\bf Proof.}
Due to Lemma~\ref{lem-BR}, there is a constant $C>0$ such that $\|[D, f(H)] \| \leq \tilde{C}_3 \lVert H \rVert\, \| [D, H]\|$. Moreover, $|f(H)| \geq c_1 \mod \Aa$ holds by functional calculus due to the normalization assumption $\sigma(H+\Aa) \cap (-1,1) = \emptyset$. 
Adding a spectral function $V=\tilde{\chi}(H^2) \in \Aa$ for $\tilde{\chi}$ a smooth positive function supported in the interval $[0,c_1^2)$ and equal to $\frac{c_1^2}{2}$ in $[0,\frac{c_1^2}{2}]$, it follows from \eqref{eq-LowBound}
\begin{align*}
(L^o_{\kappa,f})^2  
&\;\geq\; c_1^2 - \tfrac{1}{2}c_1^2 + \mu^2 \kappa^2 - \kappa \lvert[D,\tilde{H}]\rvert - \mu \kappa c_2 \, &\mod M_2(\Aa) \;\;\\
&\;\geq\; \tfrac{1}{2} c_1^2 + \mu^2 \kappa^2 - \kappa (\tilde{C}_3 \lVert H \rVert \, \| [D, H]\| + \mu c_2) &\mod M_2(\Aa)\;.
\end{align*}
Since $(L^o_{\kappa,f})^2$ is a bounded perturbation of $\kappa^2 \tilde{D}^2$ it follows that $\Aa$ is relatively $\Tt$-compact w.r.t. $(L^o_{\kappa,f})^2$. Hence, arguing as in the proof of Proposition~\ref{prop:fredholm1}, the same lower bound also holds modulo $M_2(\Kk_{\Tt})$.
\hfill $\Box$

\begin{lemma}
\label{lem-PathHomotopy}
Let $G$ be any switch function as in Proposition~\ref{prop-ExpF} with $g=1$, namely $G'$ supported in $(-1,1)$.
The straight-line paths in \eqref{eq-Path0} and \eqref{eq-Path1} are homotopic via
$$
s\in [0,1]\; \mapsto \;f_{t,s}(\lambda) \;=\; (1-s)f_{t,0}(\lambda) + s f_{t,1}(\lambda)
$$
in such a way that \eqref{eq:lowerbound} computed for $f_{s,t}$ is uniformly bounded from below by a strictly positive number for any small enough $\kappa > 0$. Moreover, the functions $s\in[0,1]\mapsto |f_{0,s}|$ and $s\in[0,1]\mapsto |f_{1,s}|$ are invertible.
\end{lemma}

\noindent{\bf Proof.}
By construction the parameter $t$ merely flips the imaginary part $$f_{t,s}=\Re e(f_{0,s}) + \frac{\imath}{2}(1-2t)\Im m( f_{0,s}).$$
We consider the homotopy for $\lvert \lambda \rvert \geq 1$ where the functions simplify to 
$$
f_{t,s}(\lambda)\;=\; \sgn(\lambda)(1+(1-s)\lvert \lambda\rvert) + \imath m (1-s) (1 - 2 t)
$$ 
and hence $c_1 \geq 1$ and $c_2 \leq 2(\|H\|+m)$ uniformly in $s,t\in[0,1]$. For $\mu$ and $\kappa$ small enough the quantity \eqref{eq:lowerbound} is therefore obviously bounded from below. Checking pointwise invertibility for $t=0$ and $t=1$ is also simple: the imaginary part never changes sign and only ever vanishes when $\lvert \lambda\rvert \geq 1$ where one always has an non-vanishing real part.
\hfill $\Box$

\vspace{.2cm}

Let us now fix $\mu$, without restriction, to the value $\mu=1$. Smallness of $\kappa$ is such that \eqref{eq:lowerbound} in Lemma~\ref{lem-PathHomotopy} holds.

\begin{corollary}
\label{coro-FirstExpression}
Let us introduce the unitary $\tilde{W}= -\imath e^{\imath \frac{\pi}{2} G(\tilde{H})}$. Then for $\kappa$ small enough 
\begin{align*}
\Tt\mbox{-}\Ind(\tilde{D}_{\kappa,\tilde{H}}) \;=\; 
-\,
\SF(\begin{pmatrix}
\kappa \tilde{D} & \tilde{W} \\ \tilde{W}^* & - \kappa \tilde{D}
\end{pmatrix},\begin{pmatrix}
\kappa \tilde{D} & \tilde{W}^*  \\  \tilde{W}  & - \kappa \tilde{D}
\end{pmatrix}).
\end{align*}
\end{corollary}

\noindent{\bf Proof.}
Start out with Lemma~\ref{lem-SFInd} and note that this straight line path there is given in \eqref{eq-Path0}. As the Fredholm property holds troughout the square $(t,s)\in[0,1]^2$ by Lemma~\ref{lemma:fredholm_localizer}, the homotopy invariance of the spectral flow as stated in Proposition~\ref{prop-SFprop}(ii) allows to deform the path \eqref{eq-Path0} into \eqref{eq-Path1} by respecting the invertibility of the end points, see Lemma~\ref{lem-PathHomotopy}.
\hfill $\Box$

\vspace{.2cm}

\noindent{\bf Proof} of Theorem~\ref{thm-Main}: As already stated above, it is sufficient to prove the equality for some $\kappa>0$ because then the constancy of the $\Tt$-index along paths of Fredholm operators allows to conclude, due to the bound \eqref{eq-kappa0bound}. Now let us start out with \eqref{eq:indtilde} and then invoke Corollary~\ref{coro-FirstExpression}:
$$
\Tt\mbox{-}\Ind({D}_{\kappa,{H}}) \;=\; 
-\,
\SF(\begin{pmatrix}
\kappa \tilde{D} & \tilde{W} \\ \tilde{W}^* & - \kappa \tilde{D}
\end{pmatrix},\begin{pmatrix}
\kappa \tilde{D} & \tilde{W}^*  \\  \tilde{W}  & - \kappa \tilde{D}
\end{pmatrix}).
$$
Set $\tilde{U} = e^{\imath \pi( G(\tilde{H})+\one)}=-e^{\imath \pi G(\tilde{H})}$. Then
by construction $\tilde{U} = \tilde{W}^2$ and therefore the adjoint action of the unitary $\mathrm{diag}(\one, \tilde{W})$ transforms the spectral flow to
\begin{align*}
\Tt\mbox{-}\Ind({D}_{\kappa,{H}})&\;=\;-\, \SF(\begin{pmatrix}
\kappa \tilde{D} & 1 \\ 1 & - \kappa \tilde{W} \tilde{D} \tilde{W}^*
\end{pmatrix},\begin{pmatrix}
\kappa \tilde{D} & \tilde{U}^*  \\  \tilde{U}  & - \kappa \tilde{W} \tilde{D}\tilde{W}^*
\end{pmatrix})
\;.
\end{align*}
The next aim is to replace $\tilde{W} \tilde{D}\tilde{W}^*$ by $\tilde{D}$ by a homotopy $s\in[0,1]\mapsto (1-s)\tilde{W} \tilde{D}\tilde{W}^* +s\tilde{D}$ leading to a homotopy of straight line paths. The difference $\tilde{W} \tilde{D} \tilde{W}^*-\tilde{D} = \tilde{W}[\tilde{D},\tilde{W}^*]$ is a bounded operator by Lemma~\ref{lem-BR}, and therefore for $\kappa$ small enough the invertibility of the end points remains valid along the homotopy as does the lower bound on the essential spectrum so that the Fredholm property is conserved throughout. Therefore
\begin{align*}
\Tt\mbox{-}\Ind({D}_{\kappa,{H}})&\;=\;\, -\SF(\begin{pmatrix}
\kappa \tilde{D} & 1 \\ 1 & - \kappa \tilde{D}
\end{pmatrix},\begin{pmatrix}
\kappa \tilde{D} & \tilde{U}^*  \\  \tilde{U}  & - \kappa \tilde{D}
\end{pmatrix})
\end{align*}
Now we are in the situation to apply Proposition~\ref{prop:indexspectralflow4} which gives
$$
 \Tt\mbox{-}\Ind({D}_{\kappa,{H}})\;=\; \,\tInd(\chi(\tilde{D} >0) \tilde{U} \chi(\tilde{D}>0) + 1 - \chi(\tilde{D}>0))$$
where we took into account that compared to the formulation of Proposition~\ref{prop:indexspectralflow4} the spectral projection is flipped, which cancels the factor of $\mbox{-}1$. The last expression  is the index pairing between $\tilde{D}$ and $\tilde{U}$ which is equal to the pairing between the undoubled Dirac operator $D$ and $\tilde{U} \ominus \one$, see \cite{CPRS3}. Since $\tilde{U} \ominus \one=U=\exp(\imath \pi( G(H)+\one))$ the expression is equal to the spectral flow $\SF_D(H)$.
\hfill $\Box$

\section{Even version}
\label{sec-Even}

A spectral triple is called even if there is a proper self-adjoint unitary $\gamma \in \Nn$ that anti-commutes with $D$, but commutes with all elements of $\mathcal{A}$. As matrices with respect to the projections $\pi_\pm = \frac{1}{2}(\gamma\pm\one)$ induced by the grading $\gamma$ one then has the decompositions
$$
D 
\;=\;
\begin{pmatrix}
0 & D_0^*\\ D_0 & 0
\end{pmatrix}, 
\qquad \sgn(D) \;=\;\begin{pmatrix}
0 & \Diracphase^*\\ \Diracphase & 0
\end{pmatrix}, 
\qquad A \;=\;\begin{pmatrix}
A_+ & 0\\ 0 & A_-
\end{pmatrix}
\;,
$$
for each $A\in \Aa$ and a partial isometry $\Diracphase$. In the following we denote $T_\pm = \pi_\pm T \pi_\pm$ for any operator where such a decomposition makes sense, and also set $\Sigma=\sgn(D)$.

\vspace{.2cm}

For even spectral triples the index pairing with any unitary vanishes, but instead there is a pairing with $K_0(\Aa)$ given by the skew-corner index \cite{CPRS3,KNR}
\begin{equation}
\label{eq-EvenIndexPairing}
\langle [P]_0, [D_0] \rangle 
\;=\; 
\Tt\mbox{-}\Ind_{P_+\cdot P_-}(P_+ \Diracphase^* P_-)
\end{equation}
where $P \in \mathscr{\Aa}^\sim$ is a projection representing the class $[P]_0-[s(P)]_0 \in K_0(\Aa)$ (and the formulas adapt to matrices in the obvious manner). 

\vspace{.2cm}

Any Callias-type operator in the sense of Definition~\ref{def-CalliasType} has a vanishing index since one has $D_{\kappa,H}=-\gamma D_{\kappa, H}^* \gamma$. To obtain an even analogue for the index theorem, let us therefore shift to non-self-adjoint potentials $T$, which form the off-diagonal part of a doubled potential $H\in M(M_2(\Aa),M_2(\Nn))$, or alternatively and more in the spirit of physical systems having an extra (so-called chiral) symmetry, the self-adjoint potential $H$ is required to be a $2\times 2$ matrix that is off-diagonal w.r.t. a natural extra grading by the third Pauli matrix $J=\diag(\one,-\one)$.

\begin{definition}
\label{def-CalliasEven}
An $\Aa$-multiplier $T\in M(\Aa,\Nn)$ is an even Callias potential if $H = \begin{pmatrix} 0 & T^* \\ T & 0
\end{pmatrix}\in M(M_2(\Aa),M_2(\Nn))$ is a Callias potential for the spectral triple $(M_2(\Nn), D\otimes \one_2, \mathscr{A}\otimes \one_2)$.
The associated even Callias-type operator is
$$
D^e_{\kappa,T} \;=\; \begin{pmatrix}
T_+ & \kappa D_0^* \\ \kappa D_0 & -T_-^*
\end{pmatrix}
\;,
$$
acting on the domain $\mathrm{Dom}(D_0)\oplus \mathrm{Dom}(D_0^*)$.
\end{definition}

Let us note the differentiability of $H$ w.r.t. $D\otimes \one$ is equivalent to the differentiability of $T$ w.r.t. $D$. The asymptotic invertibility of $H$ (contained in the notion of Callias potential) requires that there is a self-adjoint operator $V\in M_2(\Aa)$ and a $g>0$ such that $H^2+V\geq g^2 \one_2$.
Moreover, the off-diagonal nature of $J$ is equivalent to the (chiral) symmetry
$$
JHJ\;=\;-H\;,
\qquad
J\;=\;\diag(\one,-\one)
\;.
$$ 
Note that also $J=J_+\oplus J_-$ and then $J_\pm H_\pm J_\pm =-H_\pm$.  

\vspace{.2cm}

Next let us show how $D^e_{\kappa,T}$ naturally arises from the Callias operator $D_{\kappa, H} $ as given in Definition~\ref{def-CalliasType}. In fact, one readily checks
\begin{equation}
\label{eq-Deven_decomp}
D_{\kappa, H} 
\;=\; 
\begin{pmatrix}
	0 & \imath T^*_+ & \kappa D_0^* & 0 \\
	\imath T_+ & 0 & 0 &  \kappa D_0^* \\
	 \kappa D_0 & 0 & 0 & \imath T_-^* \\
	0 &  \kappa D_0 & \imath T_- & 0 
\end{pmatrix}
\;=\; 
{\Pi}^*_{\frac{3\pi}{2}}\begin{pmatrix}
	0 & -(D^e_{\kappa, T^*}) \\
	D^e_{\kappa, T} & 0
\end{pmatrix}\Pi_{\frac{3\pi}{2}}
\;, 
\end{equation}
where
\begin{equation}
\label{eq-PiPhi}
{\Pi}_\varphi 
\;=\; \begin{pmatrix}
1 & 0 & 0 & 0\\
0 & 0 & 0 & e^{\imath\varphi}\\
0 & e^{\imath\varphi} & 0 & 0\\
0 & 0 & 1 & 0
\end{pmatrix}
\;.
\end{equation}
Hence $D_{\kappa, H}$ is block off-diagonal in an appropriate basis and one of the off-diagonal entries is indeed $D^e_{\kappa,T}$, hence motivating Definition~\ref{def-CalliasEven}. The above identity also allows to deduce several analytic properties of $D^e_{\kappa,T}$ from corresponding statements for the odd case. Proposition~\ref{prop-KatoRellich} implies that $(D_{\kappa,T}^e)^*=D_{\kappa,T^*}^e$, while Proposition~\ref{prop:fredholm1} shows that $D^e_{\kappa,T}$ is a $\Tt$-Fredholm operator.  The following $K$-theoretic result now corresponds to Proposition~\ref{prop-ExpF}.

\begin{proposition}
\label{prop-IndF}
Let $T$ be an even Callias potential such that the associated $H\in M_2(M(\Aa,\Nn))$ satisfies $H^2+V > g^2\one_2$ for some $g>0$ and self-adjoint $V\in M_2(\Aa)$. For an odd switch function $G: \RM \to \RM$ as in {\rm Proposition~\ref{prop-ExpF}},  define the following self-adjoint unitary $S\in M_2(\Aa^\sim)$ and projection $P\in M_2(\Aa^\sim)$
$$
S\;=\; e^{-\imath \frac{\pi}{2}G(H)} J e^{\imath \frac{\pi}{2}G(H)}
\;,
\qquad
P 
\;=\;
\frac{1}{2}(\one_2-S) 
\;.
$$
Then the index map $\partial_1:K_1(M(\Aa,\Nn)/\Aa)\to K_0(\Aa)$ associated to the exact sequence \eqref{eq-ExactSeq} gives
$$
[P ]_0-[\mathrm{diag}(\one,0)]_0
\;=\;
\partial_0([T]_1)
\;.
$$
\end{proposition}

\noindent{\bf Proof.} This is exactly the definition of the index map.
\hfill $\Box$

\vspace{.2cm}

Note that $JHJ=-H$ implies
$$
S
\;=\; J e^{\imath \pi G(H)} 
\;=\; e^{-\imath \pi G(H)} J.
$$

\begin{definition}
\label{def-ChiralSF}
For an even Callias potential $H=-JHJ$ with off-diagonal entry $T$, the $D$-spectral flow is defined as a skew-corner index pairing \eqref{eq-EvenIndexPairing} by
$$
\SF_{D}(T) \;=\; \langle [P]_0, [D_0]\rangle \in \RM
\;,
$$
where $P$ is as in {\rm Proposition~\ref{prop-IndF}}.
\end{definition}

Now the main result of this section can be stated.

\begin{theorem}
\label{thm-MainEven}
Let $H$ be an even Callias potential with off-diagonal entry $T$ and let $\kappa_0$ be as in \eqref{eq-kappaCond}. Then for all $\kappa\in(0,\kappa_0)$ 
$$
\Tt\mbox{-}\Ind(D^e_{\kappa, T}) \;=\; \,\SF_D(T).
$$
\end{theorem}

The l.h.s. can also be understood as the supersymmetric index of the odd self-adjoint operator $\kappa (D \otimes \one_2)+ \gamma H$ (in the sense of \cite{Bunke}), though one may prefer the formulation in terms of $T$ due to the homomorphism property:

\begin{corollary}
If $T_1$,$T_2$ are even Callias potentials then $T_1 T_2$ is an even Callias potential with
$$
\Tt\mbox{-}\Ind(D^e_{\kappa, T_1 T_2}) \;=\; \Tt\mbox{-}\Ind(D^e_{\kappa, T_1})+\Tt\mbox{-}\Ind(D^e_{\kappa, T_2})
$$
for small enough $\kappa$ and therefore
$$\,\SF_D(T_1 T_2) = \,\SF_D(T_1) +\SF_D(T_2).$$
\end{corollary}
\noindent{\bf Proof.} There is a standard homotopy between $\begin{pmatrix}
	T_1 T_2 & 0 \\ 0 & 1
\end{pmatrix}$ and $\begin{pmatrix}
T_1 & 0 \\ 0 & T_2
\end{pmatrix}$ and it is not difficult to check that differentiability and asymptotic invertibility are satisfied along such a path.
\hfill $\Box$

\vspace{.2cm}

If the Dirac operator is not invertible, it is again necessary to regularize it by adding a mass term $\mu$. This can in principle be done by the usual doubling procedure \eqref{eq:doubling}, but it is more convenient to work with a unitarily equivalent representation in which the regularized Dirac operator is again off-diagonal, namely by setting
$$
\tilde{D}\;=\;\begin{pmatrix}
0 & \tilde{D}_0^* \\
\tilde{D}_0 & 0
\end{pmatrix}\;, 
\qquad
\tilde{D}_0 \;=\;
\begin{pmatrix}
\mu & -D_0^* \\ 
D_0 & \mu 
\end{pmatrix}
\;,
\qquad \tilde{\gamma}\;=\;\begin{pmatrix}
\pi_+\oplus \pi_- & 0 \\ 0 &- \pi_- \oplus \pi_+
\end{pmatrix}
\;, 
$$
and
$$
\tilde{T}\;=\;\begin{pmatrix}
\mathrm{diag}(T_+,1) & 0  \\
0 & \mathrm{diag}(T_-,1) \\
\end{pmatrix}
\;,
\qquad
\tilde{H}=\begin{pmatrix}
0 & \tilde{T}^* \\ 
\tilde{T} & 0 \\
\end{pmatrix} 
\;.
$$ 
It is then again possible to decompose $ \tilde{H}\in M_2(\Aa^\sim)$ as $\tilde{H}=\tilde{H}_+\oplus \tilde{H}_-$ by applying $\tilde{\pi}_\pm = \pi_\pm \otimes \one_2$ to each matrix entry. As before the index of the Callias operators does not depend on $\mu$ unless the mass term is too large and breaks the Fredholm property. In order to avoid clumsy notations, let us from now on simply suppose  without loss of generality that $D$ is invertible with a lower bound $|D|\geq \mu$. This also leads to some minor simplification in Lemma~\ref{lem-FredEven} below compared to Lemma~\ref{lemma:fredholm_localizer}. From now on, we thus suppress all tildes on $D$, $H$, etc. Moreover, we will assume that a scaling as in Section~\ref{sec-ProofOdd} has been carried out, assuring that $H\geq 1$ mod $\Aa$.

\vspace{.2cm}

The proof of Theorem~\ref{thm-MainEven} starts out again by applying Proposition~\ref{prop:indexspectralflow} which allows to compute the index of $D^e_{\kappa,A}$ as a spectral flow 
$$
\Tt\mbox{-}\Ind(D^e_{\kappa,T})\;=\;\SF(\begin{pmatrix}
-m & (D^e_{\kappa,T})^*\\ D^e_{\kappa,T} & m
\end{pmatrix},\begin{pmatrix}
m & (D^e_{\kappa,T})^*\\ D^e_{\kappa,T} & -m
\end{pmatrix})
\;.
$$
A permutation $\Pi_0$ defined via \eqref{eq-PiPhi} mixing the spectral eigenspaces of $\gamma$ and $J$ 
leads to
$$
\Pi^*_0
\begin{pmatrix}
m & (D^e_{\kappa,T})^*\\ D^e_{\kappa,T} & -m
\end{pmatrix}
\Pi_0
\;=\;
\begin{pmatrix}
m & T_+^* & \kappa D_0^* & 0 \\
T_+ & -m & 0 & \kappa D_0^*  \\
\kappa D_0 & 0 & -m & -T_-^*\\
0 & \kappa D_0 & -T_- & m 
\end{pmatrix} 
\;.
$$
Using $J_\pm=\diag(\one,-\one)$ as a matrix also in the mixed eigenspaces, this can be written as
$$
\Pi_0^*\begin{pmatrix}
m & (D^e_{\kappa,T})^*\\ D^e_{\kappa,T} & -m
\end{pmatrix}
\Pi_0
\;=\; \begin{pmatrix}
H_+ + m J_+ & \kappa D_0^* \otimes \one_2  \\
\kappa D_0 \otimes \one_2 & - H_- - mJ_-
\end{pmatrix}
\;,
\qquad
H_\pm
\;=\;
\begin{pmatrix}
0 & T_\pm^* \\ T_\pm & 0
\end{pmatrix}
\;.
$$
Hence by the unitary invariance of the spectral flow
\begin{equation}
\label{eq-TIntermed}
\Tt\mbox{-}\Ind(D^e_{\kappa,H})\;=\;\SF(\begin{pmatrix}
{H}_+ -m J_+ & \kappa D_0^* \otimes \one_2  \\
\kappa D_0 \otimes \one_2 & - {H}_- + mJ_-
\end{pmatrix},
\begin{pmatrix}
{H}_+ +m J_+ & \kappa D_0^* \otimes \one_2  \\
\kappa D_0 \otimes \one_2 & - {H}_- - mJ_-
\end{pmatrix})
\;.
\end{equation}
This turns out to be a better starting point for the homotopy arguments. More precisely, the operators $H_\pm+J_\pm m$ will be deformed within the set of operators of the form $f(H_\pm) +J_\pm g(H_\pm)$ to the operator $-J_\pm e^{\imath \frac{\pi}{2} G(H_\pm)}$, along a path that conserves the Fredholm property, for details see Lemma~\ref{lem-HomotopyEven} below. For that purpose, one needs a Fredholm criterion for the homotopy of paths which is the next result, a modification of Lemma~\ref{lemma:fredholm_localizer}. For a smooth odd function $f:\RM\to\RM$ and a smooth function $g:\RM\to\CM$ satisfying $g(-\lambda)=\overline{g(\lambda)}$, both compactly supported, let us introduce the associated even spectral localizer 
$$
L^e_{\kappa,f,g} \;=\;\begin{pmatrix}
f({H}_+)+J_+\, g({H}_+) & \kappa {D}_0^*\otimes \one_2 \\ 
\kappa {D}_0\otimes \one_2 & -f({H}_-)-J_-\, g({H}_-)
\end{pmatrix}
\;.
$$ 
Due to $J_\pm H_\pm J_\pm=-H_\pm$ and the symmetry of $g$, one has $(J_\pm g({H}_\pm))^*=J_\pm g({H}_\pm)$ and therefore by the Kato-Rellich theorem also $L^e_{\kappa,f,g}$ is a self-adjoint operator with domain $\mathrm{Dom}(D_0)^{\times 2}\oplus \mathrm{Dom}(D_0^*)^{\times 2}$.

\begin{lemma}
\label{lem-FredEven}
Let $T$ be an even Callias potential such that $H^2+V \geq 1$ for some $V=V^*\in M_2(\Aa)$. For $f$ and $g$ as above, associated constants 
$$
c_1^2 \;=\;\min_{|\lambda| \geq 1} 
\big(|f(\lambda)|^2+|g(\lambda)|^2\big)
\;, 
$$ 
as well as $\tilde{C}_3=\tilde{C}_3(f)+\tilde{C}_3(g)$ in terms of the constants in {\rm Lemma~\ref{lem-BR}}, suppose that $\kappa$ is such that
\begin{equation}
\label{eq:lowerbound2}
\tfrac{1}{2} c_1^2 + \mu^2 \kappa^2 - \kappa\, \tilde{C}_3 \lVert H \rVert \| [D, H]\| \;>\; 0
\;.
\end{equation}
Then $(L^e_{\kappa, f ,g})^2$ is a self-adjoint $\Tt$-Fredholm operator with spectrum mod $\Kk_{\Tt}$ bounded from below by $\frac{1}{2}c_1^2 + \mu^2 \kappa^2 - \kappa \tilde{C}_3 \lVert H \rVert \| [D, H]\|$.
\end{lemma}

\noindent {\bf Proof.}
One computes
$$
\left(L^e_{\kappa,f,g}\right)^2 
\;=\;\begin{pmatrix}
\lvert f({H}_+)\rvert^2+\lvert g({H}_+)\rvert^2 +\kappa^2D_0^*D_0 & \kappa B^*  \\ 
\kappa B & \lvert{f({H}_-)\rvert}^2+\lvert g({H}_-)\rvert^2+\kappa^2D_0 D_0^*
\end{pmatrix}
$$
with $B = D_0 (f({H}_+)+J_+g(H_+)) -(f({H}_-)+J_-g(H_-))D_0$. Noting that 
$$
\begin{pmatrix}
0 & B^* \\ B & 0
\end{pmatrix} 
\;=\; 
\big([{D}, f({H})]+J[D,g(H)]\big)\gamma
\;,
$$ 
one deduces 
$$
\left(L^e_{\kappa,f,g}\right)^2 
\;\geq\;
c_1^2+\kappa^2\mu^2-\kappa \,\tilde{C}_3 \lVert H \rVert \| [D, H]\|
\;,
$$
and can conclude the proof by the same arguments as in the proof of Lemma~\ref{lemma:fredholm_localizer}.
\hfill $\Box$

\begin{lemma}
\label{lem-HomotopyEven}
The straight-line path 
$$
t \in [0,1] \;\mapsto \;f_{t,0}({\lambda})+J g_{t,0}({\lambda}) 
\;=\; 
(1-t)({\lambda} - J m) + t ({\lambda}  + J m)
$$ 
is homotopic to the straight-line path
$$
t \in [0,1] \;\mapsto \;
f_{t,1}({\lambda}) + J g_{t,1}({\lambda}) \;=\; 
(1-t) (-J) e^{-\imath \frac{\pi}{2}G({\lambda})}
+tJ  e^{\imath \frac{\pi}{2}G({\lambda})} 
$$
via
$$
s\in [0,1]\; \mapsto \; 
f_{t,s}({\lambda})+J g_{t,s}({\lambda})  \;=\; (1-s)\big(f_{t,0}({\lambda})+J g_{t,0}({\lambda})\big) 
+ s \big(f_{t,1}({\lambda})+J g_{t,1}({\lambda})\big)
$$
in such a way that \eqref{eq:lowerbound2} computed for $f_{s,t}$ and $g_{s,t}$ is uniformly bounded from below by a strictly positive number for any small enough $\kappa > 0$. Moreover, for $t\in\{0,1\}$ the two paths $s \in [0,1] \mapsto \lvert f_{t,s}({\lambda})\rvert^2 +\lvert g_{t,s}({\lambda})\rvert^2$ are uniformly bounded away from $0$. 
\end{lemma}

\noindent{\bf Proof.} In the statement $J$ is merely used as a symbol to join the two functions $f_{s,t}$ and $g_{s,t}$.
One expands
\begin{align*}
\lvert f_{s,t}(\lambda)\rvert^2 + \lvert g_{s,t}(\lambda) \rvert^2 \;=\; 
&
\big((1-s)\lambda\big)^2 + 
\left((1-2t)(m(1-s)+s\cos(\tfrac{\pi}{2}G(\lambda)))\right)^2  
+ \left(s\sin(\tfrac{\pi}{2}G(\lambda))\right)^2\,,
\end{align*}
which upon substituting the value of $G$ for $\lvert \lambda \rvert \geq 1$ reduces to
\begin{align*}
\min_{|\lambda| \geq 1}
\big(\lvert f_{s,t}(\lambda)\rvert^2 + \lvert g_{s,t}(\lambda) \rvert^2\big) 
&\;=\;\min_{|\lambda| \geq 1}
\big( ((1-s)\lambda)^2 +\left((1-2t)m(1-s)\right)^2 + s^2 \big) 
\;\geq\; \frac{1}{2}
\;.
\end{align*}
The constant $\tilde{C}_3$  is obviously bounded by compactness. It remains to show that $\lvert f_{s,t}(\lambda)\rvert^2 + \lvert g_{s,t}(\lambda) \rvert^2$ is invertible for $t\in \{0,1\}$ and all $\lambda$. Invertibility can only fail at $\lambda=0$ since that is the only point where the first and third summand of $\lvert f_{s,t}(\lambda)\rvert^2 + \lvert g_{s,t}(\lambda) \rvert^2$ have a common zero. But then
$$
\lvert f_{s,t}(0)\rvert^2 + \lvert g_{s,t}(0) \rvert^2
\;=\; 
(1 - 2 t)^2(m(1-s)+  s)^2 
\;=\;
(m(1-s)+  s)^2
\;\geq\; 
\min\{m^2,1\}
\;,
$$
which by continuity shows the last claim. Let us stress that it is the required invertibility that effectively fixes the signs of the coefficents of $g_{t,1}$. 
\hfill $\Box$

\begin{corollary}
\label{coro-EvenIntermed}
Set $W_\pm=e^{\imath \frac{\pi}{2}G({H}_\pm)}$. Then, for $\kappa$ small enough,
$$
\Tt\mbox{-}\Ind(D^e_{\kappa,H})
\;=\;
\SF(\begin{pmatrix}
-J_+ W_+^* & \kappa D_0^* \otimes \one_2  \\
\kappa D_0 \otimes \one_2 & J_-W_-^*
\end{pmatrix},\begin{pmatrix}
J_+ W_+ & \kappa D_0^* \otimes \one_2  \\
\kappa D_0 \otimes \one_2 & -J_-W_-
\end{pmatrix})
\;.
$$
\end{corollary}

\noindent {\bf Proof.} Start out with \eqref{eq-TIntermed}, which with the notations of Lemma~\ref{lem-HomotopyEven} can be written out with diagonal entries $f_{0,0}(H_\pm)+J_\pm g_{0,0}(H_\pm)$ and $f_{1,0}(H_\pm)+J_\pm g_{1,0}(H_\pm)$. Now the straight line path can be deformed due to Lemmata~\ref{lem-FredEven} and \ref{lem-HomotopyEven}  combined with the  homotopy invariance of the spectral flow under homotopies with invertible end points. Therefore, $\Tt\mbox{-}\Ind(D^e_{\kappa,H})$ is equal to
$$
\SF(
\begin{pmatrix}
f_{0,s}(H_+)+J_+g_{0,s}(H_+) \!\!\!\!\!\!\!\!\! & \kappa D_0^* \otimes \one_2  \\
\kappa D_0 \otimes \one_2 & \!\!\!\!\!\!\!\!\!  -f_{0,s}(H_-)-J_-g_{0,s}(H_-)
\end{pmatrix},
\begin{pmatrix}
f_{1,s}(H_+)+J_+g_{1,s}(H_+) \!\!\!\!\!\!\!\!\! & \kappa D_0^* \otimes \one_2  \\
\kappa D_0 \otimes \one_2 & \!\!\!\!\!\!\!\!\! -f_{1,s}(H_-)-J_-g_{1,s}(H_-)
\end{pmatrix})
$$
for all $s\in[0,1]$. Use this for $s=1$.
As $f_{t,1}(\lambda)=0$ and $g_{t,1}(\lambda)=-(1-t) e^{-\imath \frac{\pi}{2}G({\lambda})} + t e^{\imath \frac{\pi}{2}G({\lambda})}$, replacing the definition of $W_\pm$ shows the claim.
\hfill $\Box$

\begin{lemma}
\label{lemma:tech}
For $\kappa$ small enough, 
\begin{align*}
\Tt\mbox{-}\Ind(D^e_{\kappa,H})
\;=\;
\SF(\begin{pmatrix}
0 & \kappa {D}_0 ^* \otimes \one_2  \\
\kappa {D}_0  \otimes \one_2 & 0
\end{pmatrix},\begin{pmatrix}
J_+W_+^2 & \kappa {D}_0 ^* \otimes \one_2  \\
\kappa {D}_0  \otimes \one_2 &- J_- W_-^2
\end{pmatrix})
\;.
\end{align*}
\end{lemma}

\noindent{\bf Proof.}
Let us introduce the unitary
$$
U\;=\;
e^{\imath \frac{\pi}{4} G({H})}
\;=\;
\begin{pmatrix}
U_+ & 0 \\
0 & U_-
\end{pmatrix}
\;,
\qquad
U_\pm\;=\;e^{\imath \frac{\pi}{4} G({H_\pm})} 
\;.
$$
Then $U_\pm^* J_\pm =J_\pm U_\pm$ again due to $J_\pm H_\pm J_\pm=-H_\pm$, and $U_\pm^2=W_\pm$. Applying the adjoint action of $U$ to the formula in Corollary~\ref{coro-EvenIntermed} leads to
$$
\Tt\mbox{-}\Ind(D^e_{\kappa,H})
\;=\;
\SF(\begin{pmatrix}
-J_+  & \kappa D_0^* \otimes \one_2  \\
\kappa D_0 \otimes \one_2 & J_-
\end{pmatrix},\begin{pmatrix}
J_+ W_+^2 & \kappa  U_+^* (D_0^* \otimes \one_2) U_- \\
\kappa U_-^* (D_0 \otimes \one_2) U_+ & -J_-W_-^2
\end{pmatrix})
\;.
$$
Now 
$$
\begin{pmatrix}
0 &  U_+^* (D_0^* \otimes \one_2) U_- \\
U_-^* (D_0 \otimes \one_2) U_+ & 0
\end{pmatrix}
\;=\;
U^*(D\otimes \one_2)U
\;=\;
D\otimes \one_2\,+\,U^*[D \otimes \one_2,U]
\;.
$$
As $[D \otimes \one_2 ,U]$ is bounded and then multiplied by $\kappa$, a homotopy as in the proof of Theorem~\ref{thm-Main} allows to remove the commutator so that
$$
\Tt\mbox{-}\Ind(D^e_{\kappa,H})
\;=\;
\SF(\begin{pmatrix}
-J_+  & \kappa D_0^* \otimes \one_2  \\
\kappa D_0 \otimes \one_2 & J_-
\end{pmatrix},\begin{pmatrix}
J_+ W_+^2 & \kappa  D_0^* \otimes \one_2 \\
\kappa D_0 \otimes \one_2  & -J_-W_-^2
\end{pmatrix})
\;.
$$
Finally, one can also homotopically remove the diagonal entries $\mp J_\pm$ of the left entry since these entries are required for neither the invertibility nor the Fredholm property. In fact,  as $J(-1+e^{\imath \pi G(H)}) \in M_2(\Aa)$ is relatively compact to ${D}$, the Fredholm property of all involved operators can readily be checked.
\hfill $\Box$

\vspace{.2cm}

Let us now complete the proof under the additional assumption that the spectral triple $(\Nn, D, \mathscr{A})$ is Lipshitz regular, which by definition means that $(\Nn, \abs{D}, \mathscr{A})$ is also a spectral triple. Likewise  a self-adjoint $\Aa$-multiplier is said to be Lipshitz-differentiable if it is differentiable w.r.t. both $D$ and $\abs{D}$. Once the proof is achieved for Lipshitz regular spetral triples, it will be shown in Lemma~\ref{lemma:dspectralfn} below that any spectral triple can be deformed into a Lipshitz regular one.

\vspace{.2cm}

For an even spectral triple with grading $\gamma$ consider $\pi_\pm = \frac{1}{2}(\gamma\pm\one)$ and $\Sigma = \sgn(D)$. If the triple is Lipshitz regular, then one can consider the representation $\rho: \Aa \to \Nn$ given by $\rho(a)=\pi_+ a \pi_+ + \pi_- \Sigma a \Sigma \pi_-$. Then $(\Nn, D, \rho(\Aa))$ is again an even spectral triple because
$$
[D, \rho(A)] 
\;=\; 
\pi_+ \Sigma[\abs{D},A]\pi_+ + \pi_- [\abs{D},A]\Sigma \pi_-
$$
is bounded if $[\abs{D},A]$ is bounded. The following lemma repackages a similar spectral flow argument from Section~6 of \cite{ST2021} that eventually connects to the index pairing:

\begin{lemma}
\label{lem-LipshitzFormula}
Let $(\Nn, D, \mathscr{A})$ be a Lipshitz regular even spectral triple with invertible Dirac operator. Assume that $S = \one - 2P \in \Nn$ for a projection $P = P_0 + A$ where $A \in \Aa$ with $[D,A]$ and $[\abs{D},A]$ bounded and $P_0$ a projection with $[D,P_0]=0=[\gamma,P_0]$, {\it i.e.} $P_0$ is the scalar part of $P$. Setting $\rho(S)= \one - 2(P_0 + \rho(A))$, one then has
$$
\SF(S\gamma , \rho(S)\gamma  )
\;=\; \SF(\kappa D + S\gamma, \kappa D)
$$
for $\kappa$ small enough. 
\end{lemma}

\noindent{\bf Proof.}
We use the family of approximate Dirac-operators $(D_R)_{R>0}$ of Lemma~\ref{lemma:app_tech}. For arbitrary $R>0$ we consider the norm-continuous two-parameter family 
$$
(s,t)\in [0,1]\times [0,1]\; \mapsto\; T_{s,t}\;=\; s \kappa D_R+\big((1-t) S + t\rho(S)\big)\gamma
\;.
$$
Since $(S\gamma)^2=\one=(\rho(S))^2$ and $D_R$ also anti-commutes with $\gamma$ one can write
$$T_{s,t}^2 = \kappa^2 \abs{D_R}^2 + (1-t)\kappa [D_R,S]\gamma + t\kappa [D_R,\rho(S)] + \one + 2 t(1-t) \{S, \rho(S)-S\}.$$
With the constant $c$ from Lemma~\ref{lemma:app_tech} a sufficient condition for the invertibility of the endpoints of the homotopy at $t\in \{0,1\}$ is therefore
$$\one -\kappa c (\norm{[D,S]}+\norm{[D,\rho(S)]}) > 0$$
and that is clearly the case for small enough $\kappa$. The differentiability of $S$ implies $[\Sigma,S]\in \Kk_{\Tt}$ and thus $S-\rho(S)=\pi_-\Sigma [\Sigma, S]\pi_-$ is also a $\Tt$-compact, such that then $T_{s,t} =T_{s,0} \mod \Kk_{\Tt}$ is also Fredholm for all $s,t\in [0,1]$.

In conclusion we have shown
$$
\SF(S \gamma, \rho(S) \gamma)
\;=\; 
\SF(\kappa D_R + S\gamma , \kappa D_R + \rho(S)\gamma )
\;,
$$
for arbitrary $R> 0$ and then by concatenation
$$
\SF(\kappa D_R + S\gamma , \kappa D_R + \rho(S)\gamma )
\;=\; 
\SF(\kappa D_R + S\gamma , \kappa D_R) \,+\, \SF(\kappa D_R, \kappa D_R +  \rho(S)\gamma )
\;.
$$
Finally, define the unitary $U= -\pi_- \Sigma \pi_+ + \pi_+ \Sigma \pi_-$ for which one checks that $UD_RU^*=-D_R$ and $U \rho(S)\gamma  U^* = - \rho(S)\gamma $ and hence using invariance under unitary conjugation
$$
\SF(\kappa D_R, \kappa D_R+ \rho(S)\gamma )
\;=\; 
\SF(-\kappa D_R, -(\kappa D_R +  \rho(S)\gamma )) 
\;=\; 
-\,\SF(\kappa D_R, \kappa D_R+\rho(S)\gamma ) \;=\; 0
\;.
$$
Lemma~\ref{lemma:sfcontinuous} concludes the proof since $D_R$ converges to $D$ in gap metric. 

Let us also note that formally the argument still makes sense if one directly substitutes $D$ for $D_R$, except that the homotopy above could then not be Riesz- or gap-continuous in general, as that would imply that the family at $s=0$ also has compact resolvents if $D$ has a compact resolvent (hence the approximation argument fixes a technical error in the proof of \cite[Lemma~16]{ST2021} where it was tacitly assumed that Riesz-continuity holds). 
\hfill $\Box$

\vspace{.2cm}

\noindent {\bf Proof} of Theorem~\ref{thm-MainEven}. Due to Lemma~\ref{lemma:dspectralfn} below, one may assume without loss of generality that $T$ is a Lipshitz differentiable $\Aa$-multiplier and the Dirac operator $D$  is Lipshitz regular. Then Lemma~\ref{lem-LipshitzFormula} can be applied to the expression in Lemma~\ref{lemma:tech}, by choosing $S=J e^{\imath\pi \,G({H})}=\diag(J_+W_+^2,J_-W_-^2)$ and $P_0=\binom{\one \;0}{0\;0}$. Thus 
\begin{align*}
\tInd({D}^e_{\kappa,{T}})
&\;=\; \SF(\begin{pmatrix}
0 & \kappa {D}_0 ^* \otimes \one_2  \\
\kappa {D}_0  \otimes \one_2 & 0
\end{pmatrix},\begin{pmatrix}
S_+ & \kappa {D}_0 ^* \otimes \one_2  \\
\kappa {D}_0  \otimes \one_2 & -S_-
\end{pmatrix}) \\
&
\;=\; -\SF(\begin{pmatrix}
S_+ & 0 \\ 0 & -S_-
\end{pmatrix},\begin{pmatrix}
S_+ & 0 \\ 0 & -{\Diracphase} S_+ {\Diracphase}^*
\end{pmatrix}) \\
&\;=\; \,\SF(S_-,{\Diracphase} S_+ {\Diracphase}^*)\
\end{align*}
where ${\Diracphase}$ is the phase of ${D}_0$. Recalling Proposition~\ref{prop-IndF}, Lemma~\ref{lem-SkewCorner} below now implies
$$
\tInd({D}^e_{\kappa,{T}})
\;=\; \,\tInd_{P_+\cdot P_-}(P_+ {\Diracphase}^* P_-)
\;,
$$
which according to \eqref{eq-EvenIndexPairing} and Definition~\ref{def-ChiralSF} concludes the proof.
\hfill $\Box$

\begin{lemma}
	\label{lem-SkewCorner}
	The skew-corner index can also be computed as a spectral flow via
	$$
	\tInd_{P_+\cdot P_-}(P_+ {\Diracphase}^* P_-)
	\;=\;\SF(\one-2P_-, \Diracphase (\one-2P_+) \Diracphase^*)
	\;.
	$$
\end{lemma}
\noindent {\bf Proof.}
By definition \eqref{eq-skewcorner} of the skew-corner index,
$$
\tInd_{P_+\cdot P_-}(P_+ \Diracphase^* P_-)
\;=\; 
\Tt(\Ker(P_+ \Diracphase^* P_-)\cap P_-)
\,-\,\Tt(\Ker(P_- \Diracphase P_+)\cap P_+)
\;,
$$
while the spectral flow on the r.h.s. can be computed from the Definition~\ref{def-SF2} using only the endpoints
$$
\SF(\one-2P_-, \Diracphase (\one-2P_+) \Diracphase^*)
\;=\;
\Tt\left(P_- \cap (\Diracphase (\one-P_+) \Diracphase^*)\right) - \Tt\left((\one-P_-) \cap (\Diracphase P_+ \Diracphase^*)\right)
\;,
$$
because $P_- - \Diracphase P_+ \Diracphase^* \in \Kk_{\Tt}$ follows from $[P, \Sigma]\in \Kk_{\Tt}$ which holds in any spectral triple.
Since $\Diracphase$ is unitary here, one can check that 
$$
\Ker(P_+ \Diracphase^* P_-)\cap P_- 
\;=\; 
(\Diracphase (\one- P_+) \Diracphase^*)\cap P_-
$$ 
and 
$$
\Ker(P_- \Diracphase P_+)\cap P_+ 
\;=\; 
(\Diracphase^* (\one- P_-) \Diracphase) \cap P_+ 
\;=\; 
\Diracphase^*\left( (\one-P_-)\cap (\Diracphase P_+ \Diracphase^* \right)\Diracphase
\;,
$$
such that taking traces gives the desired equality.
\hfill $\Box$


\vspace{.2cm}

To complete the proof of Theorem~\ref{thm-MainEven}, it remains to show that the Lipshitz regularity can be assumed without loss of generality. The gist of the argument, going back to a trick by Kaad \cite[Proposition 5.1]{Kaad2020}, is that by replacing the Dirac operator $D$ with $D (1+D^2)^{-\frac{r}{2}}$ for some $0 <r < 1$, one obtains an equivalent spectral triple which is Lipshitz regular. It needs to be verified that this construction is compatible with differentiability to ensure that the Callias-type operators stay Fredholm even for the regularized triple:

\begin{lemma}
\label{lemma:dspectralfn}
If $H$ is a bounded self-adjoint differentiable multiplier and $f\in\Ss^\rho(\RM)$ for $\rho < 1$, then $H$ is differentiable with respect to $f(D)$.  Assuming further $(1+f^2)^{-\frac{1}{2}} \in \Ss^{\beta}(\RM)$ for some $\beta < 0$, one has $(f(D)+\imath)^{-1} A \in \Kk_{\Tt}$ for each $A\in \Aa$ and so $(f(D), \Nn, \mathscr{A})$ is again a spectral triple.
\end{lemma}

\noindent{\bf Proof.}
For the differentiability of $H$ it is enough to verify that there is a core $\Ee \subset \mathrm{Dom}(f(D))$ for which $H\,\Ee \subset \mathrm{Dom}(f(D))$ and that $[f(D),H]$ extends from $\Ee$ to a bounded operator. One can use $\Ee=\mathrm{Dom}(D) \subset \mathrm{Dom}(f(D))$ which is preserved by $H$ by assumption. 


\vspace{.1cm}

Next let us choose a smooth switch function $\chi$ equal to $1$ on $[-1,1]$ and vanishing outside $(-2,2)$ and regularize $f_R(\lambda)=f(\lambda) \chi(\lambda R^{-1})$. There is for any $k\in \NM$ a constant $c_k$ such that $\abs{\partial^k \chi(\lambda)} \leq c_k (1+\lambda^2)^{-\frac{k}{2}}$ and then by scaling 
$$
\abs{\partial^k f_{R}} 
\;\leq\; 
\sum_{m=0}^k C_m (1+x^2)^{\frac{\rho-k}{2}}
\,\frac{1}{R^{k-m}}\, c_{k-m} 
\;\leq\; \widehat{C}_k 
(1+x^2)^{\frac{\rho-k}{2}}
$$
with constants uniformly in $R \geq 1$. With an almost analytic continuation $\tilde{f}_{R,N}$ and $\psi \in \Ee$, one can then write
\begin{align*}
[f_R(D),H]\psi
&
\;=\; \int_G (\partial_{\overline{z}} \tilde{f}_{R,N}(z)) \,[(D-z)^{-1},H]\psi \,\mathrm{d}z \wedge \mathrm{d}\overline{z}
\\
&
\;=\; 
-\int_G (\partial_{\overline{z}} \tilde{f}_{R,N}(z))\, (D-z)^{-1}[D,H](D-z)^{-1}\psi\, \mathrm{d}z \wedge \mathrm{d}\overline{z}
\end{align*}
and since $f_R(D)$ and $H$ are bounded the latter expression also holds for all $\psi \in \Hh$. Since
$$
\norm{(\partial_{\overline{z}} \tilde{f}_{R,N}(z)) (D-z)^{-1}[D,H](D-z)^{-1}}
\;\leq\; 
c_n \,\tilde{C}_{R,N+1} 
\,(1+x^2)^{\frac{\rho-1-N}{2}}
\,\lvert{\Im m (z)\rvert}^{-2+N}\,\norm{[D,H]}
$$
with $\tilde{C}_{R,N+1}$ bounded uniformly in $R$, the integral is also bounded uniformly in $R$ when substituting $N\geq 2$ and hence $\sup_{R\geq 1} \norm{[f_R(D),H]} < \infty$. For $\psi \in \Ee$ one also has $H\psi \in \Ee$ and the spectral representation shows $H f(D)=H \lim_{R\to\infty}f_R(D)\psi = \lim_{R\to\infty}H f_R(D)\psi$ and $f(D)H=\lim_{R\to\infty}f_R(D)H\psi$. Hence $[f(D),H]\psi=\lim_{R\to\infty}[f_R(D),H]\psi$ for all $\psi \in \Ee$ which implies that the commutator extends to a bounded operator. Finally $(f(D)+\imath)^{-1}A \in \Kk_{\Tt}$ for $H\in \Aa$ again follows from the functional calculus since $(f(D)+\imath)^{-1}$ can be expressed as a norm-convergent integral of terms $(D+z)^{-1}A \in \Kk_{\Tt}$.
\hfill $\Box$

\vspace{.2cm}

Assuming that the Dirac operator $D$ is invertible and let $0 < \rho < 1$ then this Lemma implies that for $D^{(\rho)} = D(1+D^2)^{-\frac{\rho}{2}}$ one has spectral triples $(\Nn,  D^{(\rho)}, \mathscr{A})$ and $(\Nn, \abs{D^{(\rho)}},\mathscr{A})$ (for the latter note that $\lambda \mapsto \abs{\lambda}(1+\lambda^2)^{-\frac{\rho}{2}}$ may be replaced by a smooth function as $0\notin \sigma(D)$). Moreover, any bounded $D$-differentiable multiplier is $D^{(\rho)}$- and $\abs{D^{(\rho)}}$-differentiable. Hence one can replace the spectral triple with a Lipshitz regular one for which $H$ is Lipshitz differentiable. 

\begin{lemma}
If $T$ is a bounded Callias potential and $(D^{(\rho)})^e_{\kappa,T}$ the even Callias-type operator obtained from pairing with the Dirac operator $D^{(\rho)}$, then  $\tInd((D^{(\rho)})^e_{\kappa,T})$ does not depend on $\rho \in [\frac{1}{2},1]$ for small enough $\kappa$.
\end{lemma} 

\noindent{\bf Proof.}
Due to the inequality $\norm{[D^{(\rho)},T]}\leq \norm{D^{-\rho}} \, \norm{[D,T]}$ one can choose $\kappa$ so small that $(D^{(\rho)})^e_{\kappa,T}$ is $\Tt$-Fredholm for all $\rho \in [\frac{1}{2},1]$ and then the result follows from homotopy invariance since the path $r \in [\frac{1}{2},1] \mapsto D^{(r)}$ is gap-continuous and $T$ a bounded perturbation.
\hfill $\Box$

\section{Callias-type operators with unbounded potentials}
\label{sec-Unbounded}

This section introduces a class of unbounded Callias potentials for which it is possible to reduce the computation of the index to the bounded case. This then allows to state and prove unbounded versions of Theorems~\ref{thm-Main} and \ref{thm-MainEven}.

\begin{definition}
An unbounded $\Aa$-multiplier $T$ is a closed operator affiliated to $\Nn$ in such a way that the bounded transform 
$$
F(T)
\;=\;
T(1+T^*T)^{-\frac{1}{2}}
$$ 
is a multiplier in $M(\Aa, \Nn)$ and $(1+T^*T)^{-\frac{1}{2}} \Aa$ is a dense subset of $\Aa$.
\end{definition}

If $\Aa$ and $\Nn$ act non-degenerately on $\Hh$, then $M(\Aa, \Nn)$ is the usual multiplier algebra and $T$ is affiliated to $\Aa$ in the $C^*$-algebraic sense of Woronowicz \cite{Woronowicz91}, however, we do not ask for that since one may want to pass to proper non-dense subalgebras of $\Aa$ later. Left multiplication by an unbounded multiplier $A\mapsto T A$ gives a closed operator from a dense subset of $\Aa$ to $\Aa$. Since functional calculus factors through the bounded transform each $C_0$-function of a self-adjoint multiplier lies in $M(\Aa,\Nn)$.

\begin{definition}
\label{def:differentiable}
A self-adjoint operator $H$ affiliated to $\Nn$ is said to be $D$-differentiable (with respect to the self-adjoint operator $D$) if there is a core $\Ee$ for $D$ such that the following holds for each $\mu \in \RM \setminus \{0\}$:
\begin{enumerate}
\item[{\rm (i)}] $(H-\imath \mu)^{-1}\Ee \subset \mathrm{Dom}(D)\cap \mathrm{Dom}(H)$ and $D(H-\imath \mu)^{-1} \Ee \subset \mathrm{Dom}(H)$.
\item[{\rm (ii)}] The operator $[D,H](H-\imath \mu)^{-1}$ extends from $\Ee$ to a bounded operator in $\Nn$.
\end{enumerate} 
\end{definition}

If the two conditions hold for some core $\Ee$, then they also hold for $\Ee=\mathrm{Dom}(D)$ \cite[Proposition~7.3]{KL0}. The dense subspace $\Dd_H = (H+\imath)^{-1}\mathrm{Dom}(D)$ is dense in $\mathrm{Dom}(D)\cap \mathrm{Dom}(H)$ w.r.t. the graph norm $\lVert \psi\rVert_{D,H}=\lVert \psi\rVert+\lVert D\psi\rVert+\lVert H\psi\rVert$. In particular, the commutator $[D,H]$ is also densely defined and symmetric on $\Dd_H $. A bounded operator $H$ is differentiable if and only if it preserves $\mathrm{Dom}(D)$ and $[D,H]$ extends to a bounded operator.

\vspace{.2cm}

The above notion of differentiability is chosen precisely such that the self-adjointness criteria from \cite{KL0} imply well-definedness of the following:

\begin{proposition}
For a differentiable $\Aa$-multiplier $H$ introduce the Callias-type operators 
$$
D_{\kappa, H} 
\;=\; 
\kappa D + \imath H
\;, 
\qquad 
D^*_{\kappa, H} 
\;=\; 
\kappa D - \imath H
\;,
$$
on the domain $\mathrm{Dom}(D)\cap \mathrm{Dom}(H)$ as well as 
$$
L_{\kappa,H} 
\;=\; 
\begin{pmatrix}
0 & D^*_{\kappa, H}\\ D_{\kappa, H} & 0
\end{pmatrix} 
\;,
$$
on the domain $(\mathrm{Dom}(D)\cap \mathrm{Dom}(H))^{\times 2}$. Then $L_{\kappa,H}$ is self-adjoint and therefore $D_{\kappa, H}$ and $D_{\kappa, H}^*$ are adjoints to each other.
\end{proposition} 

Since $D$ and $H$ are affiliated to $\Nn$ one can check from the domains that the Callias-type operators and $L_{\kappa,H}$ are affiliated to $\Nn$ and $M_2(\Nn)$ respectively (again, a closed operator $T$ is affiliated if it commutes with each unitary $U\in \Nn'$).

\begin{definition}
	An unbounded self-adjoint $\Aa$-multiplier $H$ is asymptotically invertible if there is a positive self-adjoint element $V \in \Aa$ such that $H^2+V$ has a bounded inverse (which then lies in $M(\Aa,\Nn)$). A self-adjoint $D$-differentiable $\Aa$-multiplier $H$ that is asymptotically invertible will be called an (unbounded) Callias potential.
\end{definition}

The main result is that the index theorem as stated in Theorem~\ref{thm-Main} extends to unbounded Callias potentials. For the formulation, let us note that
Proposition~\ref{prop-ExpF} remains valid if $[\chi(H + \Aa < 0)]_0 \in K_0(M(\Aa,\Nn)/\Aa)$ is replaced by $[\chi(F(H) + \Aa < 0)]_0$. In particular, the index pairing $\langle [e^{\imath\pi (G(H)+\one)}]_1, [D]\rangle \in \RM$ is well-defined.

\begin{theorem}
\label{thm-MainUnbounded}
Let $H$ be a (possibly unbounded) Callias potential for the semifinite spectral triple $(\Nn, D, \mathscr{A})$. Then there is a $\kappa_0>0$ such that for all $\kappa\in(0,\kappa_0]$,
$$
\Tt\mbox{-}\Ind(D_{\kappa, H}) \;=\; \langle [e^{\imath\pi  (G(H)+\one)}]_1, [D]\rangle
\;
$$
and $G$ any switch function chosen as in Proposition~\ref{prop-ExpF}.
More precisely, if there exists some $g>0$ and positive self-adjoint $V\in \Aa$ such that $H^2+V > g^2 \one$, one can choose 
$$
\kappa_0 
\;=\; 
\lVert [D,H](H+\imath)^{-1}\rVert^{-1} \frac{g^2}{\sqrt{1+g^2}}
\;.
$$
Such a $V$ exists for all $g>0$ if and only if the resolvent of $H$ is $\Aa$-compact, {\it i.e.} $(H+\mu \imath)^{-1} \in \Aa$, in which case  all $\kappa\in(0,\infty)$ are allowed and the resolvent of $L_\kappa$ is $\Tt$-compact.
\end{theorem}

Let us briefly discuss the last mentioned situation for the classical case of a Riemannian manifold $X$. Then $\Aa=C_0(X,\Kk(\Hh))$ and $H$ is given by an operator-valued bounded function $x\in X\mapsto H_x\in\Bb(\Hh)$. If now $x\mapsto H_x$ grows at infinity, one can indeed choose $g$ arbitrarily large and still find $V$ such that $H^2+V > g^2 \one$. This is the situation considered in the work of Robbin and Salamon \cite{RoS}.

\vspace{.2cm}

Checking the Fredholm property is more difficult in the unbounded case, since there are domain issues and also the commutator $[D,H]$ is not bounded, but only relatively $H$-bounded:

\begin{proposition}
\label{prop:fredholm}
For a Callias potential $H$ there exists some $\kappa_0 > 0$ such that $D_{\kappa,H}$ is $\Tt$-Fredholm for all $0 < \kappa \leq \kappa_0$. In particular, $\kappa_0$ can be chosen as in {\rm Theorem~\ref{thm-MainUnbounded}}.
\end{proposition}

\noindent{\bf Proof.}
As recalled above, $\Dd_H = (H+\imath)^{-1}\mathrm{Dom}(D)$ is a core for both $D_{\kappa,H}$ and $D_{\kappa,H}^*$ and contained in the domain of $[D,H]$. For $\psi\in \Dd_H$ let us consider the quadratic form
$$
\langle (\kappa D+\imath H)\psi, (\kappa D + \imath H)\psi\rangle 
\;\geq\; 
\langle H\psi,H\psi\rangle + \kappa \langle \psi, \imath[H,D]\psi\rangle
\;,
$$
which is estimated as in the proof of \cite[Lemma~7.5]{KL0}
$$
\pm\langle \psi, \imath\kappa[H,D]\psi\rangle \;\leq\; (s + \frac{\kappa^2 b^2}{4s})\langle \psi, \psi\rangle + s\langle H\psi, H\psi\rangle
$$
for any $0<s<1$.  Assume now that $H^2+V > g^2 \one$ for some positive self-adjoint $V\in \Aa$. Fixing $s=\frac{g^2}{2(1+g^2)}$ and setting $b=\lVert [D,H](H+\imath)^{-1}\rVert$ one checks that
$$
H^2 + V 
\;>\; 
g^2 \one 
\;\geq\;  
\left(\frac{s}{1-s} + \frac{\kappa^2}{4 s(1-s)}\,b^2\right)\one
$$
for all $0 \leq \kappa \leq \kappa_0 = \frac{g^2}{b\sqrt{1+g^2}}$ (which was obtained by maximizing the r.h.s. over all $0 < s <1$).

\vspace{.1cm}

Substituting that particular choice of $s$, one has
$$
\langle D_{\kappa,H}\psi, D_{\kappa,H}\psi\rangle + (1-s)\langle \psi, V\psi\rangle 
\;\geq\; 
(1-s)(\langle H\psi, H\psi\rangle 
\,+\, 
\langle \psi, V\psi\rangle - g^2 \langle \psi, \psi\rangle) 
\;>\; 0
\;,
$$ 
and hence the strict positivity of $H^2+V-g\one $ implies that $D_{\kappa,H}^*D_{\kappa,H} + (1-s)V\otimes \one_2$ is invertible. A similar argument also yields $D_{\kappa,H}D_{\kappa,H}^* + (1-s)V > 0$.

\vspace{.1cm}

Also $V$ is $\Tt$-compact relative to $L_{\kappa,H}$ and thus $L_{\kappa,H}^2$, since
$$
\tilde{V} (L_{\kappa,H}+\imath)^{-1} 
\;=\; 
\tilde{V} (D\otimes \sigma_1 +\imath)^{-1} + \tilde{V} (D\otimes \sigma_1 +\imath)^{-1} (H\otimes \sigma_1) (L_{\kappa,H}+\imath)^{-1} 
\;,
$$
where $\tilde{V} (D\otimes \sigma_1 +\imath)^{-1}\in M_2(\Kk_\Tt)$ and $(H\otimes \sigma_1) (L_{\kappa,H}+\imath)^{-1}$ since $(L_{\kappa,H}+\imath)^{-1}$ is bounded as an operator from $\Hh$ to $\mathrm{Dom}(H)$. Since $\Kk_\Tt$ is an ideal this completes the proof.
\hfill $\Box$

\vspace{.2cm}

The last statement in Theorem~\ref{thm-MainUnbounded} follows immediately from the above, since one can take for $V$ a spectral function of $H^2$ as in Proposition~\ref{prop-SmallSupp}, respectively for each function $f\in C_c(\RM)$ one can find $g$ so large that the proof implies $f(H)\in \Aa$ and $f(L_\kappa)\in M_2(\Kk_{\Tt})$. Hence the same holds for all $C_0$-functions.

\vspace{.2cm}

The dependence of $D_{\kappa,H}$ on $\kappa$ is still gap-continuous and so the index again does not depend on $\kappa$ as long as it is small enough. 
Let us now proceed to prove that the bounded transform maps unbounded Callias potentials to bounded ones with the same index. This fact immediately concludes the proof of Theorem~\ref{thm-MainUnbounded} since one can obviously write $[e^{\imath\pi  (G(H)+\one)}]_1 = [e^{\imath\pi  ((\tilde{G}\circ F)(H)+\one)}]_1$ for another switch function $\tilde{G}$ and then apply Theorem~\ref{thm-Main}. The technically most difficult part of the proof is to verify the differentiability of $F(H)$. For decaying functions the differentiability of spectral functions of $H$ again follows from the Helffer-Sj\"ostrand calculus:

\begin{lemma}
\label{lemma:hspectralfn}
If $H$ is a self-adjoint differentiable multiplier and $f \in \Ss^\rho(\RM)$ for some $\rho < 0$, then $f(H)$ preserves the domain of $D$ and $[D,f(H)]$ extends from $\mathrm{Dom}(D)$ to a bounded operator. 
\end{lemma}

\noindent{\bf Proof.}
Since $H$ is $D$-differentiable one has $(H-z)^{-1}\psi \in \mathrm{Dom}(D)$ for any $\psi \in \Ee$. Let us estimate
\begin{align*}
\lVert D(H-z)^{-1}\psi \rVert 
&\;\leq\; \lVert (H-z)^{-1}D\psi \rVert +  \lVert (H-z)^{-1}[H,D](H-z)^{-1}\psi \rVert
\\
&\;\leq\; \lvert{\Im m z\rvert}^{-1} \lVert D\psi \rVert + \lvert{\Im m (z)\rvert}^{-1}\,(1+\abs{\imath+z} \lvert{\Im m (z)\rvert}^{-1}) \, \lVert [H,D](H-\imath)^{-1}\rVert\,  \lVert \psi \rVert
\;,
\end{align*}
due to 
$$
[H,D](H-z)^{-1}
\;=\;
[H,D](H+\imath)^{-1}+[H,D](H+\imath)^{-1}(\imath+z)(H-z)^{-1}
$$
Using an extension $\tilde{f}$ of $f$ satisfying \eqref{eq:hsbounds} with $N\geq 2$ and $\rho < 0$, the integral representation \eqref{eq:hsrep} therefore also converges in the graph norm of $D$ and defines a bounded operator $f(H): \mathrm{Dom}(D)\to  \mathrm{Dom}(D)$. Using
$$
[D,f(H)] 
\;=\; 
-\int_G (\partial_{\overline{z}} \tilde{f}(z)) (H-z)^{-1}[D,H](H-z)^{-1} \mathrm{d}z \wedge \mathrm{d}\overline{z}
\;,
$$
the boundedness of the commutator follows similarly.
\hfill $\Box$

\vspace{.2cm}

The following result is morally similar to bounds obtained in \cite{CPS}, but the proof presented here avoids the use of double operator integrals.

\begin{lemma}
\label{lemma:btdifferentiable}
If $H$ is an unbounded $D$-differentiable $\Aa$-multiplier, the bounded transform $F(H)$ is also $D$-differentiable with $\norm{[D,F(H)]}\leq \norm{[D,H](H+\imath)^{-1}}$.
\end{lemma} 

\noindent{\bf Proof.} 
One needs to check that $F(H)$ maps a core of $D$ into $\mathrm{Dom}(D)$ and extends from there to a bounded operator. As recalled below Definition~\ref{def:differentiable}, $H$ is also differentiable with the core $\Ee =\mathrm{Dom}(D)$ and $\Dd_H= (H+\imath)^{-1}\mathrm{Dom}(D)$ is a core of $D$.

\vspace{.1cm}

Applying Lemma~\ref{lemma:hspectralfn} (with $\rho=-1$) to the smooth  function $F(H)(H+\imath)^{-1}$ implies $F(H) \Dd_H \subset \mathrm{Dom}(D)$. It remains to show that the commutator $[D,F(H)]$ extends from $\Dd_H$ to a bounded operator. For the commutator one has from the integral representation of fractional powers (compare \cite[Proposition 2.10]{CP}) an integral formula:
\begin{align*}
[D, F(H)]\psi 
\;=\; 
\frac{1}{\pi} \int_{0}^\infty \frac{1}{\sqrt{\lambda}} \Big(&(1+H^2+\lambda)^{-1} (1+\lambda) [D, H] (1+H^2+\lambda)^{-1}
\\
& - H(1+H^2+\lambda)^{-1} [D,H] H(1+H^2+\lambda)^{-1}\Big) \psi \,\mathrm{d}\lambda
\;.
\end{align*}
It converges absolutely in the norm of $\Hh$ for each $\psi \in \mathrm{Dom}(D)$. For the bounded self-adjoint operator $T=\imath (H-\imath)^{-1}[D,H](H+\imath)^{-1}$, one has 
$$
T^2 \;\leq\; \norm{[D,H](H+\imath)^{-1}}^2 (1+H^2)^{-1}
\;,
$$ 
and hence 
$$
T \;\leq\; \abs{T} \;\leq\; \norm{[D,H](H+\imath)^{-1}} (1+H^2)^{-\frac{1}{2}}
$$ 
by operator monotonicity of the square root. Therefore
$$
\langle \psi, \imath[D,H]\psi\rangle 
\;=\; 
\langle (H+\imath)\psi, T (H+\imath)\psi\rangle \;\leq\; \norm{[D,H](H+\imath)^{-1}} \langle \psi, (1+H^2)^{\frac{1}{2}}\psi\rangle
$$
holds for all $\psi \in \Dd_H$. On the same domain, one then has 
$$
\langle \psi, \imath [D,F(H)]\psi\rangle 
\;\leq\; 
\frac{ \lVert [D, H] (H+\imath)^{-1}\rVert}{\pi} \int_{0}^\infty \langle \psi, f_\lambda(H) \psi\rangle\, \mathrm{d}\lambda 
$$
with the positive continuous function
\begin{align*}
f_\lambda(H)
& 
\;=\;\frac{1}{\sqrt{\lambda}} 
\Big((1+H^2+\lambda)^{-1} (1+\lambda) (1+H^2)^{\frac{1}{2}} (1+H^2+\lambda)^{-1} 
\\ 
&
\;\;\;\;\;\;\;\;\;\;\;\;
+ H(1+H^2+\lambda)^{-1} (1+H^2)^{\frac{1}{2}} H(1+H^2+\lambda)^{-1}\Big)
\\
&
\;=\;\frac{\sqrt{1+H^2}}{\sqrt{\lambda}(1+H^2+\lambda)}
\;.
\end{align*}
Using the spectral measure $\mu_\psi$ w.r.t. $H$, the integral becomes
$$
\int_{0}^\infty \langle \psi, f_\lambda(H) \psi\rangle\; \mathrm{d}\lambda 
\;=\;
\int_{\RM} \int_0^\infty \frac{\sqrt{1+x^2}}{\sqrt{\lambda}(1+x^2+\lambda)}\;\mathrm{d}\lambda \,\mathrm{d}\mu_\psi(x)
\;=\;
\pi \int_{\RM} \mathrm{d}\mu_\psi(x) 
\;=\; 
\pi
\;,
$$
which shows that the commutator defines a bounded quadratic form and hence has a bounded extension with $\lVert [D, F(H)]\rVert \leq\lVert [D, H] (H+\imath)^{-1}\rVert$.
\hfill $\Box$

\begin{proposition}
\label{prop-boundedtransformodd}
If $H$ is an unbounded Callias potential, then the bounded transform $F(H)$ is also a Callias potential. Furthermore there exists a constant $\kappa_0 > 0$ such that for all $0 < \kappa \leq \kappa_0$ 
$$
\Tt\mbox{-}\Ind(D_{\kappa,F(H)})\;=\;\Tt\mbox{-}\Ind(D_{\kappa,H})
\;.
$$
\end{proposition} 
\noindent{\bf Proof.} 
By assumption there is some self-adjoint $V\in \Aa$ and $g>0$ such that $H^2 + V > g^2\one > 0$ holds. Let $\chi$ be a smooth non-increasing function which is equal to $1$ on $[0,\frac{g^2}{4})$ and vanishes outside $[0,g^2)$. We note that the proof of Proposition~\ref{prop-SmallSupp} can be adapted for unbounded $\Aa$-multipliers since the resolvents $(H^2+V-z)^{-1}$ and $(H^2-z)^{-1}$ are readily checked to lie in the norm-closed algebra $M(\Aa,\Nn)$. Hence one concludes $\tilde{V}=\chi(H^2)$ lies in $\Aa$.

\vspace{.1cm}

The spectral mapping property implies $F(H)^2 + \tilde{V} > \tilde{g}^2$  for some positive constant $\tilde{g}>0$ and therefore $F(H)$ is asymptotically invertible. Lemma~\ref{lemma:btdifferentiable} also shows that $F(H)$ is differentiable.

\vspace{.1cm}

To show the equality of indices it is enough to prove that the joint potential $H \oplus (-F(H))$ has index $0$, due to the additivity of the index. Consider the modified potential 
$$
\hat{H}_m 
\;=\;
\begin{pmatrix}
	H & m\, \tilde{V}  \\ m\, \tilde{V} & -F(H)
\end{pmatrix}
\;,
$$ 
which we check to be a differentiable self-adjoint multiplier w.r.t. $D\otimes \one_2$. That statement is clear for $m=0$ and that $\hat{H}_m$ is again a multiplier follows easily from perturbation formulas for the bounded transform (such as \cite[Lemma 2.7]{CP}). Let $\Ee=\mathrm{Dom}(D\otimes \one_2)$, then due to the domain of self-adjointness one has 
$$
(\hat{H}_m-\imath)^{-1}\Ee
\;\subset\; 
(\mathrm{Dom}(D)\cap\mathrm{Dom}(H)) \oplus \mathrm{Dom}(D) 
\;=\;
\mathrm{Dom}(D\otimes\one_2)\cap\mathrm{Dom}(\hat{H}_m)
\;,
$$ 
and using the resolvent identity to compare with $\hat{H}_0$, one also has
$$
D(\hat{H}_m-\imath \mu)^{-1}\Ee
\;\subset\; 
D(\hat{H}_0-\imath\mu)^{-1}\Ee + D(\hat{H}_0-\imath\mu)^{-1}\begin{pmatrix}
	0 & m \tilde{V}\\ m\tilde{V} & 0
\end{pmatrix}(\hat{H}_m-\imath\mu)^{-1}\Ee 
\;\subset\; 
\Ee
$$ 
since $\tilde{V}$ preserves $\mathrm{Dom}(D)$ and $(\hat{H}_m-\imath \mu)^{-1}$ preserves $\Ee$. Finally, since $[D \otimes \one_2,\hat{H}_0](\hat{H}_0-\imath \mu)^{-1}$ and $[D,\tilde{V}]$ extend to bounded operators, another application of the resolvent identity implies that $[D \otimes \one_2,\hat{H}_m](\hat{H}_m-\imath \mu)^{-1}$ extends to a bounded operator as well.

\vspace{.1cm}

Since $\hat{H}_0$ is asymptotically invertible and $\hat{H}_m-\hat{H}_0\in M_2(\Aa)$, we have shown that $\hat{H}_m$ is a Callias potential for any $m \geq 0$. From the above one also sees 
$$
\max_{m\in [0,1]} \norm{[D,\hat{H}_m](\hat{H}_m+\imath)^{-1}} \;<\; \infty 
$$ 
such that the proof of Proposition~\ref{prop:fredholm} implies that there is some $\kappa_0$ such that $D_{\kappa,\hat{H}_m}$ is $\Tt$-Fredholm for all $0<\kappa \leq \kappa_0$ and all $0 \leq m \leq 1$.

\vspace{.1cm}

For any $m \geq 0$ one can check that $\hat{H}_m$ is invertible from the square
$$
(\hat{H}_m)^2 
\;=\; 
\begin{pmatrix}
	H^2+m^2 \, \tilde{V}^2 & m(H-F(H))\tilde{V}  \\ m(H-F(H))\tilde{V}  & F(H)^2+m^2 \tilde{V}^2
\end{pmatrix}
\;,
$$
which can be diagonalized in the spectral representation. The off-diagonal part $m\tilde{V}$ does not affect the index since it is relatively $\Tt$-compact w.r.t. $D$. Fixing any $m>0$ one has $\Tt\mbox{-}\Ind(D_{\kappa,\hat{H}_m}) =0$ for small enough $\kappa$ since $D_{\kappa,\hat{H}_m}^*D_{\kappa,\hat{H}_m}$ and similarly $D_{\kappa,\hat{H}_m}D^*_{\kappa,\hat{H}_m}$ become invertible. More precisely, this follows from the proof of Proposition~\ref{prop:fredholm}, since $\hat{H}^2_m > c^2_m \one$ allows one to choose $V=0$ there. Since the index does not depend on $m$ and $\kappa$, 
$$
0\;=\;
\Tt\mbox{-}\Ind(D_{\kappa,\hat{H}_m})
\;=\;
\Tt\mbox{-}\Ind(D_{\kappa,H\oplus (-F(H)}) 
\;=\;
\Tt\mbox{-}\Ind(D_{\kappa,H})-\Tt\mbox{-}\Ind(D_{\kappa,F(H)})
\;,
$$
concluding the proof. 
\hfill $\Box$

\vspace{.2cm}

Finally let us turn to the even unbounded case.

\begin{definition}
An unbounded $\Aa$-multiplier $T$ is an even Callias potential if 
$H = \begin{pmatrix} 0 & T^* \\ T & 0
	\end{pmatrix}$ is an unbounded Callias potential for the spectral triple $(M_2(\Nn), D\otimes \one_2, \mathscr{A}\otimes \one_2)$. The associated even Callias-type operator is defined as
$$
D^e_{\kappa,T} 
\;=\; 
\begin{pmatrix}
T_+ & \kappa D_0^* \\ \kappa D_0 & -T_-^*
\end{pmatrix}
$$
on the domain $(\mathrm{Dom}(D_0)\cap \mathrm{Dom}(T_+))\oplus(\mathrm{Dom}(D_0^*)\cap \mathrm{Dom}(T_-^*))$.
\end{definition}

The unitary equivalence \eqref{eq-Deven_decomp} and the self-adjointness of $L_{\kappa,H}$ again implies that $D_{\kappa,T}^e$ is a closed affiliated operator on the stated domain and $(D_{\kappa,T}^e)^*=D_{\kappa,T^*}^e$. Now the generalization of Theorem~\ref{thm-MainEven} to unbounded potentials reads as follows:

\begin{theorem}
\label{thm-MainEvenUnbounded}
Let $T$ be an even Callias potential with $H=\begin{pmatrix} 0 & T^* \\ T & 0
\end{pmatrix}$. Then
$$
\Tt\mbox{-}\Ind(D^e_{\kappa, T}) \;=\;\, \langle [\tfrac{1}{2}(\one-J e^{\imath \pi G(H)}), [D_0]]_0\rangle 
\;
$$
for each $\kappa$ and $G$ as in {\rm Theorem~\ref{thm-MainUnbounded}}.
\end{theorem}

The proof is immediate from the bounded version and the next result.

\begin{proposition}
\label{prop-boundedtransformeven}
If $T$ is an unbounded even Callias potential, there exists $\kappa_0$ such that $D^e_{\kappa,T}$ is $\Tt$-Fredholm for all $0<\kappa\leq\kappa_0$ and $\tInd(D^e_{\kappa,T})$ does not depend on $0 < \kappa \leq \kappa_0$. Also, the bounded transform $F(T)=T(1+T^*T)^{-\frac{1}{2}}$ is a Callias-admissible potential with the same index 
$$
\tInd(D^e_{\kappa,T})
\;=\;
\tInd(D^e_{\kappa,F(T)})
\;,
$$
for small enough $\kappa$.
\end{proposition}

\noindent{\bf Proof.}
Due to the block decomposition \eqref{eq-Deven_decomp}, $D^e_{\kappa, T}$ is $\Tt$-Fredholm if $L_{\kappa, H}$ is $\Tt$-Fredholm. Hence the existence of $\kappa_0$ follows from the odd case (Proposition~\ref{prop:fredholm}).

\vspace{.1cm}

That $F(T)$ is again a Callias potential is also clear from considerations of the odd case and the fact that we defined differentiability using a doubling construction. The doubled Hamiltonian $\hat{H}_m$ from the proof of Proposition~\ref{prop-boundedtransformodd} is unitarily equivalent to an off-diagonal matrix
$$
\hat{H}_m 
\;\sim\; 
\begin{pmatrix}
	0 & \hat{T}^*_m \\ \hat{T}_m & 0
\end{pmatrix}
\;,
$$
with the operator
$$
\hat{T}_m 
\;=\; 
\begin{pmatrix}
	T & -m \chi(T^*T) \\ m \chi(TT^*) & F(T^*)
\end{pmatrix}
\;,
$$
which is therefore an even Callias-potential and invertible for any $m>0$. One argues as in the odd case that %
$$
0\;=\;
\tInd(D^e_{\kappa, \hat{T}_m})
\;=\;
\tInd(D^e_{\kappa, T\oplus (-F(T^*)})
\;=\;
\tInd(D^e_{\kappa,T})\,+\,
\tInd(D^e_{\kappa,-F(T^*)})
$$ 
for small enough $\kappa$.

\vspace{.1cm}

Replacing a Callias potential $T$ by $\lambda T$ with $\abs{\lambda}=1$ again gives a Callias potential and since $\SM^1$ is connected, $\tInd(D_{\kappa, T})=\tInd(D_{\kappa, -T})$ by homotopy. Conjugating the potential gives a factor of $-1$ and thus
$$
\tInd(D^e_{\kappa,T})
\;=\;
(-1) \,\tInd(D^e_{\kappa,-T^*})
\;=\;
\tInd(D^e_{\kappa,F(T)})
\;,
$$
concluding the proof.
\hfill $\Box$
%
%

\section{Comparison with the unbounded Kasparov product}
\label{sec-CompareKK}

This section outlines how the Callias index arises as an unbounded representative of a $KK$-group. For simplicity, it will be assumed that the Dirac operator of the semifinite spectral triple $(\mathscr{A},\Nn,D)$ is invertible (see Section~\ref{sec-ProofOdd} on how to achieve this). 


\begin{definition}
\label{def-kkcycle} 
Let $\Aa$, $\Bb$ be separable $C^*$-algebras. An unbounded Kasparov cycle $(\mathscr{A}, E, D)$ is a tuple consisting of a countably generated $\Aa$-$\Bb$-Hilbert-$C^*$-module $E$ together with an odd regular self-adjoint unbounded operator $D: \mathrm{Dom}(D)\subset E \to E$ such that	\begin{enumerate}
		\item[{\rm (i)}] $\mathscr{A} \subset \Aa$ is a dense $*$-subalgebra such that each $a \in \mathscr{A}$ preserves $\mathrm{Dom}(D)$ and the graded commutator $[D,a]$ extends to a bounded operator on $E$.
		\item[{\rm (ii)}] The products $a (D-\imath)^{-1}$ are $\Bb$-compact for all $a\in \Aa$. 
	\end{enumerate}
\end{definition}


The remainder of the section considers a semifinite spectral triple $(\mathscr{A}, \Nn, D)$ with separable $C^*$-algebra $\Aa = \overline{\mathscr{A}}$. Such a spectral triple naturally defines an unbounded or bounded Kasparov cycle if and only if $\Kk_{\Tt}$ is $\sigma$-unital. Since that condition does not generally hold in the semifinite setting one passes \cite{KNR,CareyEtAl} to the norm-closed sub-algebra $\Cc \subset \Kk_\Tt$ generated by all $*$-algebraic combinations of elements 
\begin{equation}
\label{eq-algebrac}
A [F(D),B]\,, 
\qquad [F(D),B]\,, 
\qquad F(D) A [F(D),B]
\,, 
\qquad 
\varphi(D) A
\end{equation}
with $A,B\in \Aa$, $F(D) = D(1+D^2)^{-\frac{1}{2}}$ and $\varphi \in C_0(\RM)$. Then $(\Aa, \Cc, F(D)) \in KK(\Aa,\Cc)$ defines a bounded ungraded Kasparov cycle and $\Cc$ is the smallest $C^*$-algebra for which this is the case. An important technical point is that the unbounded Kasparov cycle is constructed precisely from $D$ and $\mathscr{A}$ since only then the Callias operators can be interpreted as unbounded Kasparov products. In most cases $(\mathscr{A}, \Cc, F(D))$ is already the bounded transform of a well-defined Kasparov cycle:


\begin{proposition}	
Set $\alpha_t(A) = e^{\imath Dt} A e^{-\imath Dt}$ and let $\Aa_\alpha \subset{\Nn}$ be the smallest $\alpha$-invariant $C^*$-algebra containing $\Aa$. 
If $\Aa_\alpha$ acts non-degenerately on $\Hh$, then $(\mathscr{A}\otimes \CM_1, \Cc \otimes \CM_2, D\sigma_2)$ is an unbounded Kasparov cycle. 
\end{proposition}

\noindent{\bf Proof.}
The only non-trivial point in Definition~\ref{def-kkcycle} is the regular self-adjointness of $D$  on $\Cc$ or, equivalently, that $D$ is  affiliated to $\Cc$. To prove the latter we derive a better characterization of the algebra $\Cc$.

\vspace{.1cm}

The affiliation of $D$ to $\Nn$ implies that $\alpha$ defines a weak-$*$-continuous action on $\Nn$. Since any $A \in \mathscr{A}$ is differentiable, the identity $\alpha_t(A)-A = \int_{0}^t \imath \alpha_s([D,A])\,\mathrm{d}s$ (with convergence in the weak-$*$-topology) implies that the orbit under $\alpha$ is norm-continuous, hence $\Aa_\alpha$ is still separable and $\alpha$ extends to a strongly continuous $\RM$-action on $\Aa_\alpha$. Now define $\Cc_\alpha$ to be the separable $C^*$-subalgebra of $\Kk_{\Tt}$ spanned by the elements \eqref{eq-algebrac}, but with $\Aa$ replaced by $\Aa_\alpha$.

\vspace{.1cm}

One can next form the crossed product algebra $\Aa_\alpha\rtimes_\alpha\RM$ for which we now recall some universal properties \cite{Raeburn88}: There are embeddings $\imath_{\Aa_\alpha}: \Aa_\alpha\to M(\Aa_\alpha\rtimes_\alpha\RM)$ and $\imath_{\RM}: C_c(\RM)\to M(\Aa_\alpha\rtimes_\alpha\RM)$ such that $\Aa_\alpha\rtimes_\alpha\RM$ is generated by the linear span of $\imath_{\Aa_\alpha}(\Aa_\alpha)\imath_\RM(C_0(\RM))$. Furthermore, given a non-degenerate covariant representation $(\pi, U)$ on a Hilbert space $\Hh$, there is a non-degenerate representation $(\pi\times U): \Aa_\alpha\rtimes_\alpha\RM \to \Bb(\Hh)$ for which $(\pi\times U)\circ \imath_{\Aa_\alpha}=\pi$ and $((\pi\times U) \circ \imath_{\RM})(\varphi) = \varphi(D)$ for $D$ the self-adjoint generator of $U$. Also $\imath_\RM$ extends uniquely to $C_b(\RM)$ with the same property.

\vspace{.1cm}

The identical map $\pi: \Aa_\alpha \to \Bb(\Hh)$ and $U:\RM \to \exp(\imath D \cdot)$ trivially form a covariant representation $(\pi, U)$. By definition one has $(\pi \times U)( \Aa_\alpha \rtimes_\alpha \RM) \subset \Cc_{\alpha}$ since the latter contains the generators $\pi(a) \varphi(D)$. Moreover one has equality $(\pi \times U)( \Aa_\alpha \rtimes_\alpha \RM) = \Cc_{\alpha}$, since $[\imath_{\RM}(F), \imath_{\Aa_\alpha}(a)]\in \Aa_\alpha \rtimes_\alpha\RM$ holds for any smooth switch function like $F$ (see {\it e.g.} \cite{Lesch91}).

\vspace{.1cm}

From the above one concludes that a dense subset of $\Cc_{\alpha}$ is given by all elements of the form 
$$
\sum_{k=1}^N \varphi_k(D) \alpha_{t_k}(A_k)
\;=\;
\sum_{k=1}^N \varphi_k(D) e^{\imath D t_k} \,A_k e^{-\imath D t_k}
$$
with $\varphi_1,...,\varphi_N \in C_c(\RM)$ and $A_1,...,A_N\in \Aa$. Therefore $D$ is densely defined on $\Cc_\alpha$ and clearly $(D+\imath)^{-1}\Cc_\alpha\subset \Cc_{\alpha}$ is also a norm-dense subset. Hence $D$ is affiliated to $\Cc_\alpha$ and regular self-adjoint in the Hilbert-module sense.

\vspace{.1cm}

To complete the proof let us show that $\Cc_{\alpha}=\Cc$, for which it is only necessary to verify $\varphi(D) A e^{-\imath D t} \in \Cc$ for all $\varphi\in C_c(\RM)$, $t\in \RM$, $A \in \Aa$. Choose approximate units $(B_n)_{n\in \NM}$ for $\Aa$ and $(\Psi_m)_{m\in \NM}$ for $C_0(\RM)$. Then 
$$
\varphi(D) A e^{-\imath D t} 
\;=\; 
\lim_{n\to\infty} \varphi(D) A B_n e^{-\imath D t}
\;=\; 
\lim_{n\to\infty} \lim_{m\to\infty} \varphi(D) 
A B_n e^{-\imath D t} \Psi_m(D)
\;,
$$
converges in norm, since the $B_n$ and $\Psi_m(D)$ are also approximate units for $(\pi\times U)(\Aa_\alpha \rtimes_\alpha \RM)$. That shows that $\Cc$ is norm-dense in $\Cc_{\alpha}$. \hfill $\Box$

\vspace{.2cm}

Let us now compare our main result for unbounded Callias operators to approaches using the unbounded Kasparov product (specifically \cite{vD,KL} which treat the classical case). Semifinite spectral triples often arise in noncommutative geometry from an unbounded Kasparov cycle $(\mathscr{A}\otimes \CM_1, E_{\Bb} \otimes \CM_2, D\sigma_2) \in KK^{-1}(\Aa, \Bb)$ where $\Bb$ is a separable $C^*$-algebra that carries a densely defined faithful lower semi-continuous trace $\Tt$, $E_{\Bb}$ a countably generated right $\Bb$-module and $D$ a regular self-adjoint unbounded operator on $E_{\Bb}$. Both $\mathscr{A}$ and $\Bb$ act naturally on the Hilbert space $\Hh$ obtained by completing the submodule of $E_{\Bb}$ for which $\Tt_\Bb(\langle e,e\rangle_{E_{\Bb}})<\infty$ in the obvious norm. In that situation one obtains a semifinite spectral triple $(\Nn, D, \mathscr{A})$ with $\Nn = \Bb''$ to which $\Tt_\Bb$ extends as a normal semifinite faithful trace $\Tt$. All examples described in Section~\ref{sec-Examples} below can be written in that form for some natural algebra $\Bb$. If $\Aa$ acts non-degenerately, it is completely general since, as shown above, for any semifinite spectral triple the minimal choice $\Bb=\Cc=E_{\Bb}$ is available.

\vspace{.2cm}

If the potential $H$ is a self-adjoint unbounded $\Aa$-multiplier with resolvent in $\Aa$, then it defines an odd unbounded Kasparov cycle $(\CM, \Aa \otimes \CM_1, H \sigma_2) \in KK^1(\CM, \Aa)$. The Callias operator $H\sigma_1 + D\sigma_2$ on $E_\Bb$ should then represent the product class $[H\sigma_2]\otimes_{\Aa\otimes\CM_1} [D\sigma_1] \in KK(\CM, \Bb) \simeq K_0(\Bb)$, given some compatibility conditions between $H$ and $D$ which are similar to the differentiability that we impose here (a possible set of conditions may be derived from {\it e.g.} \cite[Theorem 7.4]{LM2019}). 
Composing the product class with the homomorphism $\Tt_*: K_0(\Bb)\to \CM$ computes the $\Tt$-index, while applying $\Tt_*$ to the product of the bounded transforms $[F(H\sigma_1)]\otimes_{\Aa\otimes \CM_1} [F(D\sigma_2)]$ recovers the index pairing $\langle [e^{\imath \pi F(H)}]_1, [D]\rangle$ (see \cite{CareyEtAl}). Since the bounded and unbounded picture of $KK$-theory are isomorphic \cite{Kaad2020,vDM}, one concludes that Theorem~\ref{thm-MainUnbounded} holds in that special case. For the case of a spectral triple over $\Bb(\Hh)$, a more detailed proof can also be found in \cite{BKR2020}, though compared to our notations the regularity assumptions are formulated in terms of the Cayley transform of $H$, which is another unitary representing the class $[e^{\imath \pi F(H)}]_1$. The even case can be handled similarly with the product represented by the self-adjoint operator $D + \gamma H$ (compare \cite{Bunke}).

\vspace{.2cm}

The $KK$-theoretic approach has certain advantages, in particular, the class of the Callias operator may carry finer topological invariants besides the numerical index and also the associativity of the Kasparov product can then be further applied to prove more specialized formulas for the index. A severe limitation is that apparently the potential $H$ must always be unbounded with $\Aa$-compact resolvent, {\it i.e.} $(H+\imath)^{-1}\in \Aa$ which is a stronger condition than our asymptotic invertibility. That is necessary for the obvious cycle to define a class $KK^1(\CM,\Aa)$ in unbounded $KK$-theory, though there might be more complicated constructions to handle a bounded potential that is invertible up to $\Aa$. In the classical commutative case of potentials on a manifold, one can already treat the larger class of asymptotically invertible potentials with pointwise compact resolvents by amplifying them with a growing function on the underlying manifold (as it is done in \cite{vD,KL}). 

\section{Examples}
\label{sec-Examples}

\subsection{The classical Callias index theorem}
\label{sec-ClassCall}

The first example, which was already discussed briefly above is the classical geometric situation with Callias-type operators on a Riemannian manifold $X$. We will consider a possibly infinite dimensional vector bundle $E$ over $X$ with typical fiber isomorphic to a Hilbert space $\Hh_0$. For $D$ a weakly elliptic first-order differential operator we have a spectral triple $\big(\Bb(L^2(X, \Hh_0)), D, C^\infty_c(X)\otimes \KM(\Hh_0)\big)$. If $H=(H_x)_{x\in X}$ is a self-adjoint fibered operator that is differentiable (with derivative $(\nabla H_x)_{x\in X}$ bounded relative to $H$) and invertible modulo $C_0(X)\otimes \KM(\Hh_0)$ then the technical conditions of our index theorem can all be verified and it reproduces the result of Kaad and Lesch \cite{KL} which used the Kasparov product. If $\Hh_0$ is finite-dimensional and $H$ invertible outside a compact set $K$, there are other ways to compute the index, for example, the original index theorem by Callias \cite{Cal} on $\RM^n$ expresses the index as an integral over the boundary of a large enough sphere and more generally the index theorem by Anghel \cite{Ang2} gives a similar generalization to manifolds with warped ends. In both of these situations one only needs to know the potential $H$ on a lower-dimensional submanifold that envelops all singular points of the potential. Explicitly, in the case $X=\RM^n$, $n$ odd, with $D$ the Euclidean Dirac operator one has
\begin{equation}
\label{eq-boundaryindex}
\Ind(\kappa D + \imath H) \;\sim\; \int_{\partial B_R(0)} \Tr((Q \mathrm{d}Q)^{\wedge (n-1)}) 
\end{equation}
for $Q= H \lvert{H\rvert}^{-1}$ invertible outside $B_R(0)$. In contrast, our index formula with standard index computations would give a volume integral:
\begin{equation}
\label{eq-bulkindex}
\Ind(\kappa D +  \imath H) \;\sim\; \int_{\RM^n} \Tr((U^* \mathrm{d}U)^{\wedge n}) 
\;,
\end{equation}
with $U=e^{\imath\pi (G(H)+\one)}$ or some other representative with sufficiently fast decaying derivatives. 

\vspace{.2cm}

There is a simple $K$-theoretical relation between those formulations. Let us assume that $H$ is not only invertible up to $\Aa=C_0(X)$, but already up to $\Aa'=C_0(K)$ for $K\subset X$ a compact set. Since $\Aa'$ is an ideal in $\Aa$, one then also has $H\in M(\Aa', \Nn)$ and so the exponential map in $K$-theory gives an element of $K_1(C_0(K))$. One can do even better since there is a commutative diagram
$$
\begin{tikzcd}
0 \arrow[r] &  C_0(K) \arrow[d, "\rho"] \arrow[r] & C_b(X) \arrow[d, "\rho"] \arrow[r,"q"] & C_b(X)/C_0(K) \arrow[d, "r"] \arrow[r] & 0 \\
0 \arrow[r] &  C_0(K^\circ) \arrow[r] & C(K) \arrow[r,"q"] & C(\partial K) \arrow[r] & 0
\end{tikzcd}
$$
where $r$ is the restriction and $\rho$ acts as the identity on $C_0(K)$. Hence the naturality of the exponential map implies 
$$
\partial_0 [\chi(H + \Aa < 0)]_0 
\;=\; 
\partial_0 [\chi(r(H) < 0)]_0
\;.
$$ 
This shows that, as expected, the spectral flow $\SF_D(H)$ naturally depends only on the class that $r(H)=H\rvert_{\partial K}$ defines in $K_0(C(\partial K))$.  For a more detailed examination of those $K$-theoretical aspects of Callias-type operators we refer to the work of Bunke \cite{Bunke}. 
The equality between the expressions \eqref{eq-bulkindex} and $\eqref{eq-boundaryindex}$ can be derived from the boundary maps of $K$-homology \cite{BDT89}, which are dual to the $K$-theoretic connecting maps, however, in a more general setting constructing an explicit representative for the odd $K$-homology class on $C(\partial K)$ that computes the same pairing as the spectral flow can involve subtle geometric and analytic issues. 

\vspace{.2cm} 

In that sense our main index theorem on Callias-type operators contains only part of the information of the original theorem by Callias. It is an interesting problem to find additional analytic and algebraic data from which one can canonically construct a spectral triple for the boundary class also in the noncommutative case. Partial solutions may be provided by the construction of relative spectral triples \cite{FGMR19}.


\vspace{.2cm} 

Another natural question is whether it is possible to reduce the index computation to a compact hypersurface also in the case of an infinite-dimensional vector bundle. In general the answer is negative, though, since due to Kuipers' theorem any potential $H$ that is invertible on $\partial K$ may be homotopic to another potential $\tilde{H}$ that has the same index but is flat in the sense that $\tilde{H}\rvert_{\partial K} = \one - 2P_0$ with $P_0 \in \Bb(\Hh_0)$ some fixed projection. Hence the non-trivial index is invisible on $\partial K$. One way to evade such counterexamples is to consider a more restricted class of Callias-type operators, for example, those of the form $H=H_0 + V$ with $H_0$ a fixed self-adjoint operator with compact resolvent and $V: X \to \Bb(\Hh_0)$ a norm-continuous family of self-adjoint operators. In that case again one can compute the index from a $K_1$-class over $C^*(H_0)+C_0(K)\otimes \KM$ obtained through the exponential map from a $K_0$-class over $C^*(H_0)\rvert_{\partial K}+C(\partial K)\otimes \KM$. It must therefore be possible to compute the index using only the potential on the boundary, though we are not aware of any known formulas.

\vspace{.2cm} 

Finally, let us note an interpretation as spectral flow under the additional assumption $H_x$ is invertible for all but finitely many isolated points $x_0,...,x_N\in X$. In that case \eqref{eq-boundaryindex} decomposes into a sum of contributions of small spheres around each $x_i$, each of which is individually integer-valued. This is analogous to the way that usual spectral flow counts the number of eigenvalue crossings. In physics one uses such expressions to assign topological charges to stable band-touching points and the spectral flow is therefore the total charge. A particular set-up of this type is considered in \cite{Kub} where the potential $H$ is a linear combination of Clifford algebra generators with essentially commuting coefficients. If the critical points of $H$ are not isolated or $\Hh_0$ is infinite-dimensional the spectral flow is more difficult to interpret. For even dimensions our analogue of Section~\ref{sec-Even} can be used to define topological charges for potentials satisfing a chiral symmetry of the type $JHJ=-H$.

\subsection{The Boutet de Monvel index theorem}

The framework of Section~\ref{sec-ClassCall} can be transposed to an even $2n$-dimensional complete Riemannian manifold $X$. Let then $E$ be a possibly infinite dimensional vector bundle over $X$ with typical fiber Hilbert space $\Hh_0$. Furthermore there is supposed to exist an involution $\gamma=(\gamma_x)_{x\in X}:E\to E$ and a weakly elliptic first-order differential operator $D$ satisfying $\gamma D\gamma=-D$ and which results in an even spectral triple $\big(\Bb(L^2(X, \Hh_0)), D, C^\infty_c(X)\otimes \KM(\Hh_0)\big)$. The even Callias potential is then given by a self-adjoint multiplication operator $H=(H_x)_{x\in X}$ on $E\oplus E$ satisfying $\gamma H\gamma=H$, $JHJ=-H$ with $J=\diag(\one_\Ee,-\one_\Ee)$ and being invertible modulo $C_0(X)\otimes \KM(\Hh_0)$. In the grading of $\gamma$ one then has $H=H_+\oplus H_-$ and both $H_\pm$ are off-diagonal in the grading of $J$ with lower left entry $T_\pm$. 

\vspace{.2cm}

The set-up of the index theorem of Boutet de Monvel \cite{Bou,GH} assumes that $X$ is a strongly pseudoconvex domain in $\CM^{n}$ equipped with the Bergmann metric, $E$ is finite-dimensional and given by the differential forms on $X$ of type $(n,p)$ graded by the parity of $p$ and then $D$ is the Dolbeaut operator.
Furthermore $H_+=H_-$ is supposed to extend smoothly to the boundary $\partial X$. A related setting is obtained by simply choosing $X=\RM^{2n}$ with the euclidean metric and $D$ the associated Dirac operator on $E=\RM^{2n}\otimes \CM^{N}$ where $\CM^N$ is the representation space of the Clifford algebra with $N$ generators. Then the index theorem states \cite{Bou,GH,Bunke}
$$
\Tt\mbox{-}\Ind(D^e_{\kappa, T})
\;\sim\;
\int_{\partial X} \Tr(J(Q \mathrm{d}Q)^{\wedge (2n-1)}) 
\;,
$$
where again $Q=H|H|^{-1}$ and in the case $X=\RM^{2n}$ one replaces $\partial X$ by $\partial B_R(0)$ with $R$ sufficiently large. The r.h.s. is an odd Chern number in the representation given, {\it e.g.} \cite{CSB}. Theorems~\ref{thm-MainEven} and \ref{thm-MainEvenUnbounded} connect $\Tt\mbox{-}\Ind(D^e_{\kappa, T})$ to an integral over $X$ rather than its boundary, and hence do not directly provide the right side. However, in the case of finite-dimensional fibers one can again argue as in Section~\ref{sec-ClassCall}. To determine conditions under which such a formula holds also with infinite dimensional fibers remains an open problem.

\subsection{A generalized Robbin-Salamon theorem}

As a first non-commutative example we consider a generalized Robbin-Salamon theorem. Let $\Cc$ be a separable $C^*$-algebra with a densely defined faithful lower semicontinuous trace $\tau$ and a strongly continuous automorphic $\RM$-action $\alpha$ that leaves $\tau$ invariant (everything below directly transposes to $\RM^n$-actions). Then the crossed product algebra $\Aa = \Cc \rtimes_\alpha \RM$ has an induced dual trace $\Tt$ and a dual $\RM$-action $\hat{\alpha}$. Let further $\Mm$ be the von Neumann algebra generated by $\Cc$ in the semicyclic GNS representation for $\Tt$ and $\Hh$ the corresponding representation space. The regular representation $\pi$ of $\Aa$ acts on the Hilbert space $L^2(\RM, \Hh)$ such that $\alpha$ is generated by right translation, {\it i.e.} $\pi \circ \alpha_t = \mathrm{Ad}(U_t) \circ \pi$ with $U_t = e^{ t \partial}$. Then $\Tt$ extends to a semifinite normal faithful trace on the von Neumann algebra $\Nn=\pi(\Aa)''$. Let $\mathscr{A}$ be the dense $*$-subalgebra of elements $A\in \Aa$ that can be written in the form $\pi(A)=\int_{\RM} \pi(f(t))U_t \mathrm{d}t$ with a function $f: \RM \to \mathrm{Dom}(\tau)\subset \Aa$ that is smooth and rapidly decaying in the norm $\lVert A \rVert + \tau(\lvert A\rvert)$ and $D=-\imath \partial$ then $(\Nn, D, \mathscr{A})$ is a semifinite spectral triple (compare {\it e.g.} \cite{CareyEtAl}). If $H$ is a differentiable self-adjoint multiplier invertible modulo $\Aa$, then the index theorem implies that
$$
\hat{\Tt}\mbox{-}\Ind(\kappa D + H) \;=\; \SF_D(H)
\;,
$$
and by choosing a representative $U\in \one + \mathscr{A}$ of the class $[e^{\imath \pi (G(H)+\one)}]_1$, one can compute the spectral flow \cite{Phi}
\begin{equation}
\label{eq:ncwinding}
\SF_D(H) \;=\; \langle [U]_1, [D]\rangle \;=\; {\Tt}((\one-U^*)[\partial, U])\;.
\end{equation}
The r.h.s. is the non-commutative winding number as is expected for an analytic formula for spectral flow. The appropriate setting for the theorem in \cite{RoS}  is a trivial action $\alpha$ in which case $\Aa = C_0(\RM, \Cc)$ consists of paths in the von Neumann algebra $\Mm = \pi(\Cc)''$. Since $\tau$-traceclass elements are dense in $\Cc$ one has $\Cc \subset \Kk_{\tau}$ and hence invertibility of $H \in M(\Aa) \subset C_b(\RM, \Mm)$ modulo $\Aa$ means that $H$ is a continuous path of $\tau$-Fredholm operators. In that case, the r.h.s. of \eqref{eq:ncwinding} computes the usual semifinite spectral flow, in fact, it is almost exactly the definition of spectral flow for gap-continuous paths (see Appendix~\ref{app-SemiFolw}, note however that for an unbounded $H$ to be a multiplier here, it must describe a Riesz-continuous path).

\subsection{Index theorems for topological insulators}

Let us now discuss a more complicated non-commutative example coming from the theory of topological insulators \cite{PS}. To keep the discussion simple we consider a two-dimensional example with magnetic field, but no disordered potential. Thus the observable algebra is the two-dimensional non-commutative torus $\mathfrak{a}_\theta$ with twisting angle $\theta$ generated by two unitaries with the commutation relation $v_1 v_2 = e^{\imath \theta} v_2 v_1$. Let $c_*(\ZM)$ be the algebra of sequences which admit limits for $\pm \infty$ and $c_0(\ZM)$ the subalgebra for which those limits vanish. Let then $\widehat{\Aa}$ be the $C^*$-algebra generated by $c_*(\ZM)$ and the unitaries $v_1$,$v_2$ with the additional commutation relations $fv_1 = (f\circ T_1) v_1$ and $fv_2 = v_2 f$ with $T_1: c_*(\ZM) \to c_*(\ZM)$ left translation. Each element $a\in \widehat{\Aa}$ has a representation as a formal sum $a=\sum_{x,y\in \ZM} f_{x,y} v_1^x v_2^y$. Consider the ideal $\Aa \subset \widehat{\Aa}$ of those elements for which the coefficient functions are in $c_0(\ZM)$. Then one has an exact sequence
$$
0\, \to\, \Aa\, \to\, \widehat{\Aa} \,\to\, \mathfrak{a}_\theta \oplus \mathfrak{a}_\theta \,\to\, 0
$$
obtained by evaluation at $\pm \infty$. On $\mathfrak{a}_\theta$ one can introduce a finite trace $\tau$ and on $\Aa$ a densely defined lower semi-continuous trace $\hat{\tau}$ such that 
$$
\tau(v_1^x v_2^y) \;=\; \delta_{x,0}\, \delta_{y,0}\;, 
\qquad 
\hat{\tau}(f v_1^x v_2^y) \;=\; \delta_{x,0}\, \delta_{y,0} \sum_{k\in \ZM} f(k)
\;,
$$
holds for all $x,y\in \ZM$ and $f\in \ell^1(\ZM)$. Let $\mathscr{A}$ be the subalgebra of all $\sum_{x,y\in \ZM} f_{x,y} v_1^x v_2^y \in \Aa$ for which $\lvert f_{x,y}(k)\rvert$ decays faster than any inverse polynomial in $x,y,k$. One can represent $\widehat{\Aa}$ on the Hilbert space $\ell^2(\ZM^2)$ in such a way that $c_*(\ZM)$ acts by multiplication and $v_1$,$v_2$ as magnetic shifts. Furthermore there are the two position operators $X_1$,$X_2$ acting on the standard basis of $\ell^2(\ZM^2)$ by $X_i e_x=x_i e_x$. The commutators $[X_i, \cdot]$ produce densely defined derivations on $\Aa$ and $\mathfrak{a}_\theta$. As the Dirac operator, let us use $D=X_2$ which results in a spectral triple $(\Bb(\ell^2(\ZM, \CM^2)), D, \mathscr{A})$. One can therefore consider Callias-type operator with potentials $H$ in the multiplier algebra $M(\Aa)$. Any such multiplier has a representation $H = \sum_{x,y\in \ZM} h_{x,y}v_1^x v_2^y$ with coefficient functions $h_{x,y}\in\ell^\infty(\ZM)$. Assume that $\lVert h_{x,y}\rVert_\infty$ decays faster than any inverse polynomial in $x,y$ from which one can check that a $H$ is bounded and differentiable in our sense. If $H$ is invertible modulo $\Aa$, then by Theorem~\ref{thm-Main} its index is given by 
$$
\Ind(\kappa D +\imath H) \;=\; \SF_D(H)
\;=\;
\langle [U]_1, [X_2] \rangle
\;,
$$
with $U = e^{\imath \pi (G(H)+\one)} \in \one + \Aa$. In fact, one has $U\in \one+\mathscr{A}$ and this index can be computed explicitly \cite{PR}
$$ 
\langle [U]_1, [X_2] \rangle
\;=\; 
\hat{\tau}((\one-U^*)[X_2,U])
\;.
$$
In physics the algebra $\mathfrak{a}_\theta$ describes an observable algebra for two-dimensional tight-binding models which are invariant under magnetic translations, while $M(\Aa)$ more generally allows modulations with respect to the $x_1$-direction. In particular, a self-adjoint multiplier $H\in \widehat{\Aa} \subset M(\Aa)$ represents the Hamiltonian for a system with an interface at the line $x_1=0$ between two asymptotic "bulk" Hamiltonians $H_\pm \in \mathfrak{a}_\theta$ which describe the local Hamiltonian far away from the interface for $x_1\to \pm \infty$. The number $\SF_D(H)$ again makes sense as a non-commutative spectral flow: if the flow  along the "path" $H$ connecting the invertible Hamiltonians $H_+$, $H_-$ is non-trivial, then $H$ itself cannot be invertible ("the spectral gap closes") and this fact only depends on $H$ up to homotopy. 

\vspace{.2cm}

The known results on the bulk-boundary correspondence of such operators form a close analogue of the Callias-index formula. Indeed, for $H\in \widehat{\Aa}$ the class $[U]_1\in K_1(\Aa)$ of the spectral flow is the image under the exponential map of the class in $[P_+\oplus P_-] \in K_0(\mathfrak{a}_\theta\oplus \mathfrak{a}_\theta)$ of the Fermi projections $P_\pm = \chi(H_+ < 0)$. It is then known ({\it e.g.} \cite{KSV}) that 
\begin{align*}
\langle \partial_0[P]_0, [X_2] \rangle 
&
\;=\; 
\langle [P]_0, [X_1\otimes \sigma_1 + X_2\otimes \sigma_2]\rangle 
\\
&
\;=\; \imath \,\tau(P_+[[X_1,P_+][X_2,P_+]])
\;-\;\imath\, \tau(P_-[[X_1,P_-][X,P_-]])
\end{align*}
which shows that the index can be computed from the boundaries at $\pm \infty$. Compared to the Callias index formula the situation seems inverted since the boundary now actually represents a higher dimensional space. This is not too unusual since cyclic cohomology is $2$-periodic and so dualities can affect the apparent dimensions of cocycles and algebras. 

\vspace{.2cm}

This example can be generalized in different ways. The noncommutative torus does not really play a role in the arguments. One can construct analogous spectral triples and bulk-boundary sequences for any twisted crossed product $\Cc \rtimes \RM^d$ or $\Cc \rtimes \ZM^d$ where the base algebra $\Cc$ admits a densely defined faithful lower semicontinuous trace. When one generalizes, this naturally leads to semifinite spectral triples, {\it e.g.} if one chooses $\Cc=C(K)$ for some compact metric space $K$ with measure $\mu$, the spectral triple will be based on the type-I-von Neumann algebra $L^\infty(K,\mu)\otimes \Bb(\ell^2(\ZM^2))$ with trace $\int_K \mathrm{d}\mu\otimes \Tt$. For higher dimension $d>2$ one can also construct spectral triples with Dirac operators that involve less than $d-1$ spatial directions, so that the spectral triple is then naturally based on a type $\mathrm{II}_\infty$-von Neumann algebra. Examples for multipliers $H$ with non-trivial indices can then be given in terms of Hamiltonians of so-called weak topological insulators, see \cite{ST2021}.

\appendix

\section{Semifinite index}
\label{app-SemiInd}

Let $\Nn$ be a semifinite von Neumann algebra with a normal semifinite faithful trace $\Tt$. This appendix briefly reviews the theory of semifinite index and its continuity properties \cite{Ta,BCP,Lesch04}.  The domain of $\Tt$ is by definition given by $\Nn_\Tt=\{A\in \Nn: \Tt(\abs{A})<\infty\}$. It is a $*$-algebra and an ideal in $\Nn$. It becomes a Banach-$*$-algebra when supplied with the submultiplicative norm $\norm{A}_{\Tt}=\norm{A}+\Tt(\abs{A})$ and its $C^*$-completion $\Kk_\Tt \subset \Nn$ is the algebra of $\Tt$-compact operators. The quotient $\Nn/\Kk_\Tt$ is called the Calkin algebra and the quotient map will be denoted by $\pi$. The $\Tt$-essential spectrum of $A\in \Nn$ is then defined by $\sigma_{\mathrm{ess}}(A)=\sigma(\pi(A))=\sigma(A+\Kk_\Tt)$. 

\vspace{.2cm}

A possibly unbounded operator $T$ affiliated to $\Nn$ is called $\Tt$-Fredholm if there is a continuous function $\chi:[0,\infty)\to[0,1]$ with $\chi(0)=1$  such that $\chi(T^*T)\in\Kk_\Tt$ and $\chi(TT^*)\in\Kk_\Tt$. 
This is equivalent to the existence of a pair of operators $K,K'\in\Kk_\Tt$ such that $T^*T+K$ and $TT^*+K'$ are invertible.
The set of $\Tt$-Fredholm operators will be denoted $\Ff(\Nn)$ and intersection with the self-adjoint operators by $\Fsa(\Nn)$. 

\vspace{.2cm}

If $T \in \Ff(\Nn)$, then also $T^* \in \Ff(\Nn)$ and furthermore the kernel projection $\Ker(T)$ lies in $\Kk_\Tt$. It is important to note that any $\Tt$-compact projection is automatically $\Tt$-finite and therefore one has a well-defined index
$$
\tInd(T)
\;=\; 
\Tt(\Ker(T))-\Tt(\Ker(T^*)) \; \in\; \RM
\;.
$$
The index is invariant under addition of $\Kk_{\Tt}$ perturbations and constant on norm-connected components of $\Ff(\Nn)\cap \Nn$. 

\vspace{.2cm}

An element $T \in P\Nn Q$ for two projections $P,Q \in \Nn$ is called $P\cdot Q$-Fredholm if $T^*T$ and $TT^*$ are $\Tt$-Fredholm in the corner algebras $Q\Nn Q$ and $P\Nn P$ respectively. One then defines more generally the skew-corner index \cite{CPRS2} by
\begin{equation}
\label{eq-skewcorner}
\tInd_{P \cdot Q}(T) 
\;= \;
\Tt(\Ker(T) \cap Q) - \Tt(\Ker(T^*) \cap P)
\;.
\end{equation}

For the study of the spectral flow of unbounded Fredholm operators in Appendix~\ref{app-SemiFolw}, one needs to introduce topologies on $\Fsa(\Nn)$, of which there are several distinct ones \cite{Lesch04,Wahl2008}. The most important ones are the Riesz-topology and the gap topology, induced respectively by the metrics
$$
{d}_R\,:\, \Ff_{\mathrm{sa}}\times \Ff_{\mathrm{sa}}\,\to\,\RM_\geq\,, 
\qquad
d_R(T_1,T_2) \;=\; \lVert F(T_1)-F(T_2)\rVert
\;,
$$
where $F$ is the bounded transform and
$$
{d}_G\,:\, \Ff_{\mathrm{sa}}\times \Ff_{\mathrm{sa}}\,\to\,\RM_\geq\,, 
\qquad
d_G(T_1,T_2) \;=\; \lVert \Cc(T_1)-\Cc(T_2)\rVert \;=\; \lVert (T_1+\imath)^{-1} - (T_2+\imath)^{-1}\rVert
\;,
$$
with $\Cc(T)=(T-\imath)(T+\imath)^{-1}$ the Cayley transform. A sequence of self-adjoint operators converges in the gap topology if and only if it converges in the norm-resolvent sense. Hence if $(T_t)_{t\in [0,1]}$ is a path in $\Fsa(\Nn)$ is gap-continuous, then $(f(T_t))_{t\in [0,1]}$ is therefore a norm-continuous path in $\Nn$ for any $f\in C_0(\RM)$. The gap topology is weaker than the Riesz topology since it does not imply continuity under the bounded transform.

\vspace{.2cm}

%

By the theorem of Cordes-Labrousse, all those topologies are equivalent to the norm-topology when restricting to bounded self-adjoint operators. The gap topology can be extended to non-self-adjoint Fredholm operators by setting 
$$
\tilde{d}_G(T_1,T_2)
\;=\;d_G(\begin{pmatrix}
0 & T_1^* \\ T_1 & 0
\end{pmatrix},\begin{pmatrix}
0 & T_2^* \\ T_2 & 0
\end{pmatrix})
\;,
$$ 
and the $\Tt$-index of unbounded $\Tt$-Fredholm operators stays constant under gap-continuous homotopies \cite[Proposition 5.2]{Wahl2008}.

\section{Semifinite spectral flow}
\label{app-SemiFolw}

This appendix recalls the definition and properties of the spectral flow in a semifinte von Neumann algebra. For paths of bounded self-adjoint operators this is reviewed in \cite{BCP}. For paths of unbounded self-adjoint operators, depending on the notion of continuity there are different possible ways to define spectral flow  which we now describe in some detail since it is relevant for the main part of this article. Spectral flow for gap-continuous paths using the notion of a non-commutative winding number has been introduced in the Hilbert space setting by \cite{BLP} and extended to the semifinite setting by \cite{Wahl2008}.

\vspace{.2cm}

Consider the Banach $*$-algebra of differentiable paths
$C^1_0([0,1], \Nn_\Tt)$ with norm 
$$
\norm{f}_C
\;=\;
\sup_{t\in[0,1]}\norm{f(t)}_\Tt \,+\, \norm{f'(t)}_\Tt
\;,
$$ 
which is dense in the $C^*$-algebra $C_0([0,1],\Kk_\Tt)$ with spectrally invariant inclusion (the latter follows from the inequality $\norm{f g}_C \leq \norm{f}\,\norm{g}_C+\,\norm{f}_C\,\norm{g}$ via a standard argument using geometric series, see also \cite{Schweitzer}). The noncommutative winding number defined by
$$
\mathrm{wind}_\Tt\,:\, C^1_0([0,1],\Nn_\Tt)\times C^1_0([0,1], \Nn_\Tt)\,\to\, \CM\;, 
\qquad 
\mathrm{wind}(f_1,f_2)
\;=\;
\imath \int_0^1 \Tt(f_1(t) f'_2(t))
\;,
$$
is a cyclic $1$-cocycle and therefore pairs with odd $K$-theory groups. Due to spectral invariance, one has $K_1(C^1_0([0,1],\Nn_\Tt))) \simeq K_1(C_0([0,1], \Kk_\Tt))$ and more strongly any class  $[f]_1\in K_1(C_0([0,1], \Kk_\Tt))$ defined by a unitary path $f \in \one + C_0([0,1], \Kk_{\Tt})$ can be represented by a unitary path $\tilde{f}\in \one + C^1_0([0,1], \Nn_{\Tt})$ such that the real-valued pairing 
$$
\langle [f]_1, \mathrm{wind}_\Tt\rangle 
\;=\; 
\mathrm{wind}_{\Tt}(\tilde{f}^*-\one, \tilde{f}-\one)
$$
is well-defined and does not depend on the choice of representative.

\begin{definition}[\cite{Wahl2008}]
\label{def-SF}
Let $t\in [0,1] \mapsto T_t$ be a gap-continuous path in $\Fsa(\Nn)$ with invertible endpoints. 
One can always choose a so-called switch function $G:\RM \to \RM$ for the path, which is a smooth function with $\mathrm{supp}(G') \subset (-1,1)$, $G(\pm 1)=\pm 1$ and  $\one - G(T_t)^2\in \Kk_{\Tt}$ for all $t$. Then the norm-continuous unitary path $t\in [0,1]\mapsto e^{\imath \pi (G(T_t)+\one)}$ lies in $\one + \Kk_{\Tt}$ and the spectral flow
$$
\SF(\{T_t\}_{t\in[0,1]}) 
\;= \;
\langle [e^{\imath \pi (G(T)+\one)}]_1, \mathrm{wind}_\Tt\rangle
\;
$$
is well-defined and does not depend on the choice of $G$.
\end{definition}

The important technical point here is that $e^{\imath \pi G}-1$ is a continuous compactly supported $C_0(\RM)$-function which vanishes on the $\Tt$-essential spectrum. Therefore gap-continuity is sufficient, but in exchange the endpoints have to be invertible.

\vspace{.2cm}

For Riesz-continuous paths the spectral flow can also be defined more directly as the flow of spectrum from the negative to the positive: 

\begin{definition}[\cite{BCP}]
\label{def-SF2}
For projections $P,Q\in\Nn$ with $\norm{\pi(P-Q)}<1$ define the essential codimension
$$
\ec (P,Q)
\;=\;
\Tt\big((1-P)\cap Q\big)\,-\,\Tt\big((1-Q)\cap P\big)
\;.
$$
For a  Riesz-continuous path  $t\in [0,1] \mapsto T_t$ in $\Fsa(\Nn)$ one can always choose a partition $0=t_0 < t_1 < ... < t_{K+1} = 1$ such that 
$$
\norm{\pi\left(\chi(T_{s} \geq 0)-\chi(T_{t} \geq 0)\right)} 
\;\leq\; 
\tfrac{1}{2},
\;
\qquad \forall s,t\in [t_{k},t_{k+1}]
$$ 
holds for all $k=0,\ldots,K$. In that case the spectral flow is given by
$$
\SF(\{T_t\}_{t\in[0,1]}) 
\;= \;
\sum_{k=0}^{K} \ec\left(\chi(T_{t_k} \geq 0),\chi(T_{t_{k+1}} \geq 0)\right)
\;.
$$
\end{definition}

If the endpoints of the Riesz-continuous path are invertible both notions coincide. Therefore the spectral flow for gap-continuous paths is well-defined by choosing for each endpoint a $\Tt$-compact perturbation $Q_0, Q_1$ such that $T_i+Q_i$ is invertible and to set 
\begin{align*}
\SF(\{T_t\}_{t\in[0,1]}) 
\;=\, &\; \SF\big(\{T_t + (1-t)Q_0 + t Q_1 \}_{t\in[0,1]}\big) 
\;+\; 
\SF\big(\{T_0 + t Q_0\}_{t\in[0,1]}\big)
\\
&
\; +\; \SF\big(\{ T_1 + (1-t) Q_1\}_{t\in[0,1]}\big)
\;,
\end{align*}
where the second two terms use the definition for Riesz-continuous paths and only the former the one for gap-continuous ones. One has the following properties:

\begin{proposition}
\label{prop-SFprop}
Let  $t\in [0,1] \mapsto T_t  $ and $t\in [0,1] \mapsto T'_t$ be gap-continuous paths in $\Fsa(\Nn)$.
\begin{itemize}
\item[{\rm (i)}]{\rm (Triviality)} If $T_t$ has a bounded inverse for each $t\in [0,1]$ then  $\SF(\{T_t\}_{t\in[0,1]})=0$.
\item[{\rm (ii)}] {\rm (Homotopy invariance)} If the two paths are connected by a gap-continuous (respectively Riesz-continuous) homotopy $(t,s)\in [0,1]\times [0,1] \mapsto T_{s,t}$ within $\Fsa(\Nn)$ with $T_{0,t}=T_t$, $T_{1,t}=T'_t$ and such that the endpoints $T_{s,0}$ and $T_{s,1}$ are invertible for each $s\in [0,1]$, then 
$$
\SF(\{T_t\}_{t\in[0,1]}) \;=\; \SF(\{T'_t\}_{t\in[0,1]})
\;.
$$
\item[{\rm (iii)}] {\rm (Concatenation)} If $T_1 = T'_0$, then 
$$
\SF(\{T_{t}\}_{t\in[0,1]} * \{T'_{t}\}_{t\in[0,1]}) \;=\; \SF(\{T_{t}\}_{t\in[0,1]})\;+\;\SF(\{T'_{t}\}_{t\in[0,1]})
\;,
$$
with $*$ denoting concatenation of paths.
\item[{\rm (iv)}] {\rm (Homomorphism)}  
$$
\SF(\{T_t \oplus T'_t\}_{t\in[0,1]}) \;=\; \SF(\{T_{t}\}_{t\in[0,1]})\;+\;\SF(\{T'_{t}\}_{t\in[0,1]})
\;.
$$
\end{itemize}
\end{proposition}

For straight-line paths the spectral flow will be abbreviated by 
$$
\SF(T_0,T_1)
\;=\;
\SF(\{(1-t)T_0+t T_1\}_{t\in[0,1]})
\;.
$$ 

Let us now recall some relations between spectral flow and the $\Tt$-Fredholm index:

 \begin{proposition}[Proposition 5.1 \cite{Wahl2008}]
 \label{prop:indexspectralflow}
For $T$ a possibly unbounded $\Tt$-Fredholm operator and any $m>0$,
$$
\Tt\mbox{-}\Ind(T) 
\;=\; 
\SF(\begin{pmatrix}
 	-m & T^* \\ T & m
\end{pmatrix},\begin{pmatrix}
 	m & T^* \\ T & -m
\end{pmatrix})
\;.
$$
\end{proposition}

Then there are more specific spectral flow formulas for unitary conjugates \cite[Theorem 4.2]{CGPR}

\begin{theorem}
\label{theo:indexspectralflow2}
Let $D$ be a self-adjoint invertible $\Tt$-Fredholm operator affiliated to $\Nn$. If $U\in \Nn$ is a unitary that preserves $\mathrm{Dom}(D)$, $[D,U]$ extends to a bounded operator in $\Nn$ and such that $(D+\imath)^{-1}(U-\one) \in \Kk_{\Tt}$ and $(D+\imath)^{-1}[D,U] \in \Kk_{\Tt}$, then
$$\tInd(PUP+\one-P) \;=\; \SF(U^*DU, D)$$
where $P = \chi(D>0)$.
\end{theorem}

The bounded version of this formula is \cite[Section 5]{BCP}

\begin{proposition}
\label{prop:indexspectralflow3}
If $T$ is a self-adjoint involution and $U\in \Nn$ a unitary with $[T,U]\in \Kk_{\Tt}$, then
$$\tInd(PUP+\one-P) \;=\; \SF(T, U^*TU)$$
where $P=\chi(T<0)$.
\end{proposition}

Theorem~\ref{theo:indexspectralflow2} requires that $[D,U]$ must be relatively $D$-compact for the path to be Fredholm. While such a condition often is satisfied in applications, it is sometimes inconvenient {\it e.g.}  in the setting of spectral triples without smoothness assumptions. We therefore provide an alternative:

\begin{proposition}
\label{prop:indexspectralflow4}
Let $D$ be a self-adjoint invertible $\Tt$-Fredholm operator affiliated to $\Nn$ and $U \in \Nn$ a unitary that preserves $\mathrm{Dom}(D)$, $[D,U]$ extends to a bounded operator in $\Nn$ and $(U-\one)(D+\imath)^{-1},(U^*-\one)(D+\imath)^{-1} \in \Kk_{\Tt}$. Set $P=\chi(D<0)$, then
$$\tInd(PUP + \one - P) \;=\; \SF(\begin{pmatrix}
\kappa D & \one \\ \one & -\kappa D
\end{pmatrix},\begin{pmatrix}
\kappa D & U^* \\ U & -\kappa D
\end{pmatrix})
$$
holds for all $\kappa>0$ so small that $\kappa \norm{[D,U]} < 1$.
\end{proposition}

The proof starts out with a technical lemma:

\begin{lemma}
\label{lemma:app_tech}
Let $D$ be an unbounded self-adjoint invertible operator and $H$ a bounded self-adjoint operator which preserves $\mathrm{Dom}(D)$ and for which $[D,H]$ extends to a bounded operator. Choose an even smooth function $g:\RM \to [0,1]$ supported in $[-2,2]$ and equal to $1$ on $[-1,1]$. Set $\chi_R=g(R^{-1} D)$ and define a net $(D_R)_{R>0}$ of bounded self-adjoint operators by
$$
D_R \;=\; D \chi_R \,+\, R (1-\chi_R) \sgn(D)
\;.
$$
Then $D_R$ converges to $D$ w.r.t. the gap metric for $R\to \infty$ and there exists a universal constant $c>0$ such that 
\begin{equation}
\label{eq:drcomm}
\norm{[D_R, H]} \;\leq \;c \norm{[D,H]}
\end{equation}
independent of $D$, $R$ and $H$.
\end{lemma}

\noindent{\bf Proof.}
The convergence is readily seen in the spectral representation. For \eqref{eq:drcomm} let us first recall the bound \cite[Lemma 10.15]{GVF}
$$
\norm{[f(D),H]}
\;\leq\; 
(2\pi)^{-1} \left(\int_{\RM} \abs{t} \hat{f}(t) dt\right)\norm{[D,H]} 
\;=\; 
(2\pi)^{-1} \lVert{\widehat{f'}}\rVert_{L^1(\RM)}\,\norm{[D,H]}
$$
applicable to smooth functions $f\in C^\infty_c(\RM)$ where $\hat{f}$ denotes the Fourier transform. By multiplying with a smooth approximate unit of the Fourier algebra $\Ff^{-1}L^1(\RM)$ the bound generalizes to functions $f$ without compact support, but for which $f' \in C_c^\infty(\RM)$. Since one can write $D_R=f(R^{-1}D)$ for such a function $f \in C^\infty(\RM)$ a scaling argument therefore shows that \eqref{eq:drcomm} holds with $c=(2\pi)^{-1} \lVert{\widehat{f'}}\rVert_{L^1(\RM)}$.
\hfill $\Box$

\vspace{.2cm}

\noindent{\bf Proof (of Proposition~\ref{prop:indexspectralflow4}).}
It is sufficient to prove the result for some $\kappa > 0$ that is as small as necessary, since one can then increase $\kappa$ up to the stated value using homotopy invariance.

\vspace{.1cm}

By a standard argument $[P,U]\in \Kk_{\Tt}$ \cite{GVF} and thus Proposition~\ref{prop:indexspectralflow3} and additivity imply
$$\tInd(PUP + \one - P)  \;=\; \SF(\begin{pmatrix}
\kappa (\one-2P) & 0 \\ 0 & -\kappa(\one-2P)
\end{pmatrix},\begin{pmatrix}
\kappa U^*(\one-2P)U & 0 \\ 0 & -\kappa (\one-2P)
\end{pmatrix})\,.
$$
The endpoints of the path are invertible and introducing an off-diagonal constant term only increases the spectral gap, thus
$$\tInd(PUP + \one - P)  \;=\; \SF(\begin{pmatrix}
\kappa (\one-2P)& \one \\ \one & -\kappa (\one-2P)
\end{pmatrix},\begin{pmatrix}
\kappa U^*(\one-2P)U & U^* \\ U & -\kappa(\one-2P)
\end{pmatrix})\,.
$$
To check the Fredholm property along that straight-line homotopy one notes
$$\begin{pmatrix}
\kappa (\one-2P)-\kappa t U[P,U^*]& s(t + (1-t)U^*) \\ s(t+(1-t)U) & -\kappa (\one-2P)
\end{pmatrix}^2 \;\geq\; \kappa^2 + s^2(1-2t)^2  \mod \Kk_{\Tt}$$
The right endpoint at $(s,t)=(1,1)$ has a spectral gap in the interval $[-1,1]$ and assuming $\kappa < \frac{1}{4}$, the gap is not closed if one replaces $U^*(\one-2P)U= (\one-2P)- 2U^*[P,U]$ by $\one-2P$ using an additive perturbation with norm $\norm{[P,U]}\leq 2$. That compact perturbation also does not affect the Fredholm properties, therefore
$$
\tInd(PUP + \one - P)  \;=\; \SF(\begin{pmatrix}
\kappa (\one-2P)& \one \\ \one & -\kappa (\one-2P)
\end{pmatrix},\begin{pmatrix}
\kappa (\one-2P) & U^* \\ U & -\kappa(\one-2P)
\end{pmatrix}).
$$
For arbitrary $R>0$ we use the approximation $D_R$ of Lemma~\ref{lemma:app_tech} and consider the norm-continuous homotopy
$$(\gamma,t) \in [0,1]\times [0,1] 
\;\mapsto\; T_{\gamma,t}
\;=\;
\begin{pmatrix}
	\kappa D_R \abs{D_R}^{\gamma-1} & t\one +(1-t)U^* \\ t\one +(1-t)U & -\kappa D_R \abs{D_R}^{\gamma-1}
\end{pmatrix}.
$$

We must show that all $T_{\gamma,t}$ are Fredholm with invertible endpoints at $t\in\{0,1\}$ for some small enough $\kappa$. At the left endpoint
$$T_{\gamma,0}^2 \;=\; \begin{pmatrix}
	\kappa D_R \abs{D_R}^{\gamma-1} & \one \\ \one & -\kappa D_R \abs{D_R}^{\gamma-1}
\end{pmatrix}^2 \;=\; (\kappa^2 \abs{D_R}^{2\gamma} +\one)\otimes \one_2$$
and at the right
$$T_{\gamma,1}^2 \;=\; \begin{pmatrix}
	\kappa D_R \abs{D_R}^{\gamma-1} & U^* \\ U & -\kappa D_R \abs{D_R}^{\gamma-1}
\end{pmatrix}^2 \;=\; (\kappa^2 \abs{D_R}^{2\gamma} +\one- \kappa c \norm{D^{-1}}^{\gamma-1} \norm{[D,U]}) \one_2\;, $$
with the constant $c$ from \eqref{eq:drcomm} and where we used a known estimate for the commutator with fractional powers \cite[(10.58)]{GVF} 
$$\norm{\left[D_R\abs{D_R}^{\gamma-1}, U\right]} 
\;\leq \;
\norm{\abs{D_R}^{\gamma-1}} \norm{[D_R,U]} 
\;\leq\; c \norm{D^{-1}}^{\gamma-1} \norm{[D,U]}
\;.$$

Hence invertibility holds if we assume that $\kappa$ is small enough. The relative compactness further implies 
$$
[f(D),U]\;=\;f(D)(U-\one)-(U-\one)f(D)\in \Kk_{\Tt}
$$ 
for any function $f\in C_c(\RM)$ and since also $[P,U] \in \Kk_{\Tt}$ one concludes  $\left[D_R \abs{D_R}^{\gamma-1}, U\right] \in \Kk_{\Tt}$ for all $\gamma \in [0,1]$. Computing $T_{\gamma,t}^2$ one therefore finds
\begin{align*}
\begin{pmatrix}
	\kappa D_R \abs{D_R}^{\gamma-1} & t\one + (1-t)U^* \\ t\one + (1-t)U & -\kappa D_R \abs{D_R}^{\gamma-1}
\end{pmatrix}^2  \; &= \; (\kappa^2 \abs{D_R}^{2\gamma}+ 1 + t(1-t)(U + U^*-2))\one_2 \;\;\;\mod \Kk_{\Tt} \\
\; &\geq \; \kappa^2  \min(\norm{{D^{-1}}}^{-2\gamma}, R^{2\gamma}) \one_2 \;\;\;\mod \Kk_{\Tt}
\end{align*}
where we used $\norm{t(1-t)(U + U^*-2)}\leq 1$. Thus $T_{\gamma,t}$ is Fredholm for all $\gamma,t \in [0,1]$ and homotopy invariance implies
$$
\tInd(PUP + \one - P)  \;=\; \SF(\begin{pmatrix}
	\kappa D_R& \one \\ \one & -\kappa D_R
\end{pmatrix},\begin{pmatrix}
	\kappa D_R & U^* \\ U & -\kappa D_R
\end{pmatrix})
$$
for all $R > 0$ and some fixed $\kappa > 0$. The proof is then completed by taking the limit $R\to \infty$ as the following Lemma shows.
\hfill $\Box$

\vspace{.2cm}

\begin{lemma}
\label{lemma:sfcontinuous}
\begin{enumerate}
\item[{\rm (i)}] Let  $(T_n)_{n\in \NM}$ be a sequence of gap-continuous paths $T_n = (T_{n,t})_{t\in [0,1]}$ in $\Ff_{\mathrm{sa}}(\Nn)$ converging uniformly in gap metric to some path $T$, {\it i.e.} 
\begin{equation}
\label{eq:cont}
\lim_{n\to\infty} \sup_{t\in [0,1]} d_G(T_{n,t}, T_t) \;=\; 0
\end{equation}
If the endpoints of $T_{n}$,  $T$ are invertible and $\abs{T_{n,t}} > g\one \mod \Kk_{\Tt}$ holds for all $t\in [0,1]$, $n\in \NM$ with some fixed constant $g>0$ then the spectral flow is continuous
$$\SF(\{T_t\}_{t\in[0,1]})\;=\;\lim_{n\to\infty}\SF(\{T_{n,t}\}_{t\in[0,1]})\;.$$
\item[{\rm (ii)}] The convergence condition \eqref{eq:cont} holds in particular for paths of the form $T_{n,t}=D_n + H_t$ with $(D_n)_{n\in \NM}$ sequence of self-adjoint operators affiliated to $\Nn$ that converges to $D$ w.r.t. the gap metric and $H$ a norm-continuous path in $\Nn_{\mathrm{sa}}$.
\end{enumerate}
\end{lemma}

\noindent {\bf Proof.}
For {\rm (ii)} we note the resolvent identity
\begin{align*}
(D + &H_t + \imath)^{-1}-(D_n + H_t + \imath)^{-1} \\
&=\;\left(1-(D+H_t+\imath)^{-1} H_t\right) \left((D + \imath)^{-1}-(D_n + \imath)^{-1}\right) \left(1-H_t(D_n+H_t+\imath)^{-1} \right) 
\end{align*}
which implies $d_G(T_n,T) \leq (1+\norm{H})^2 \, d_G(D_n,D)$. Similarly, one estimates 
$$
\norm{(T+z)^{-1} - (S+z)^{-1}} 
\;\leq\; 
\left(1+\frac{\abs{\imath-z}}{\abs{\Im m ( z)}}\right)^2 d_G(T,S)
$$ 
for all $z\in \CM\setminus \RM$ and hence the Helffer-Sjostrand calculus may be used to show that the map $T \in \Ff_{\mathrm{sa}}(\Nn) \mapsto f(T) \in \Nn$ is uniformly continuous for each fixed $f \in C_c^\infty(\RM)$ in the sense that
\begin{equation}
\label{eq:cont2}
\norm{f(T)-f(S)}
\;\leq\; 
C_f \, d_G(T,S)
\;.
\end{equation}

By assumption on $T_n$ there is a gap in the $\Tt$-essential spectrum which is independent of $n$ and $t$. We may therefore also choose the normalizing function $G$ in Definition~\ref{def-SF} to be independent of those parameters. Combining \eqref{eq:cont} and \eqref{eq:cont2}  shows
$$
\lim_{n\to\infty} \sup_{t\in[0,1]} \norm{e^{\imath \pi(G(T_{n,t})+\one)}-e^{\imath \pi(G(T_{t})+\one)}}
\;=\;
0
\;,
$$
{\it i.e.} the unitary path determining the spectral flow is norm-convergent. Hence $[e^{\imath \pi(G(T_{n})+\one)}]_1$ is eventually constant with limit $[e^{\imath \pi(G(T)+\one)}]_1$, which implies
$$
\SF(\{T_t\}_{t\in[0,1]}) 
\;= \;
\langle [e^{\imath \pi (G(T)+\one)}]_1, \mathrm{wind}_\Tt\rangle
\; = \; \lim_{n\to\infty}
\langle [e^{\imath \pi (G(T_n)+\one)}]_1, \mathrm{wind}_\Tt\rangle
\; = \; \lim_{n\to\infty}
\SF(\{T_{n,t}\}_{t\in[0,1]})
\;,
$$
so that the proof is concluded.
\hfill $\Box$

\vspace{.2cm}

\noindent {\bf Acknowledgements:} T.~S. received funding from the {\it Studienstiftung des deutschen Volkes}. This work was also supported by the DFG grant SCHU 1358/6-2. 


\end{document}